\documentclass[a4paper,11pt]{article}
\pdfoutput=1 %

\usepackage{jheppub} %
\usepackage[T1]{fontenc} %
\usepackage{enumitem}

\usepackage{mathrsfs} %
\usepackage{calligra} %
\usepackage{braket} %
\usepackage{slashed} %
\usepackage{cancel}
\usepackage{soul}

\usepackage{tikz}
\usetikzlibrary{positioning, calc}
\usetikzlibrary{decorations.markings}

\newcommand{\badat}{\begin{alignedat}}
\newcommand{\eadat}{\end{alignedat}}

\newcommand{\p}{{\partial}}
\newcommand{\zb}{{\bar z}}

\newcommand{\ep}{{\epsilon}}

\newcommand{\scri}{{\mathcal I}}

\newcommand{\calA}{{\mathcal A}}
\newcommand{\calD}{{\mathcal D}}
\newcommand{\calF}{{\mathcal F}}

\newcommand{\Y}{{\mathcal Y}}

\title{
The Classical Super-Phaserotation Infrared Triangle\\
{\small Classical Logarithmic Soft Theorem as Conservation Law in (scalar) QED}}
\author[a]{Sangmin Choi,}\emailAdd{s.choi@uva.nl}
\author[b]{Alok Laddha,}\emailAdd{aladdha@cmi.ac.in}
\author[a]{Andrea Puhm}\emailAdd{a.puhm@uva.nl}

\affiliation[a]{Institute for Theoretical Physics, University of Amsterdam, PO Box 94485, 1090 GL Amsterdam, The Netherlands}
\affiliation[b]{Chennai Mathematical Institute, H1, SIPCOT IT Park, Siruseri, Kelambakkam 603103, India}
\abstract{
The universality of the logarithmic soft photon theorem in four dimensions can be traced to an infinite-dimensional asymptotic symmetry which acts as a local phase rotation on matter as we have shown in \cite{Choi:2024ygx}.
Here we extend our earlier results for the charges associated to these superphaserotations to all orders in the coupling and prove that their conservation is exactly the classical logarithmic soft photon theorem discovered by Saha, Sahoo and Sen \cite{Saha:2019tub}.
We furthermore generalize the formulae for the associated electromagnetic displacement memory and its tail from particles to scalar matter fields. This completes the classical superphaserotation infrared triangle. 
}
\begin{document} 
\maketitle
\section{Introduction and summary}

A detailed understanding of asymptotic symmetries in gauge theories and gravity has been a topic of resurgent interest over the past decade. These symmetries imply universal soft factorization theorems that are common to all theories with the same gauge group but with arbitrary matter content and generic interactions. An example is the set of theories with $U(1)$ gauge group where the photon is the massless gauge boson. In all such theories, scattering processes involving a low-energy photon obey Weinberg's soft photon theorem whose universality (theory-independence) is a consequence of an (infinity) of conservation laws associated to {\it superphaserotations}\footnote{In the literature these asymptotic symmetries are usually referred to as ``large gauge transformations''. To avoid any association with gauge redundancies we will instead refer to them as superphaserotations since they act on charged matter fields as local rotations of the phase.} with an angle-dependent symmetry parameter. 

Over the past five years, thanks to the efforts of Sahoo,  Sen and their collaborators \cite{Laddha:2018myi, Laddha:2018vbn, Sahoo:2018lxl, Saha:2019tub, Sahoo:2021ctw, Krishna:2023fxg}, a new understanding of universal factorisation theorems in effective field theories with $U(1)$ gauge group and in gravitational effective theories has emerged. This class of factorisation theorems are known as {\it logarithmic soft theorems}. When an amplitude is expanded in the energy of the gauge boson, beyond the leading Weinberg pole the soft factors have logarithmic dependence in the energy. Unlike the subleading soft theorems of tree-level amplitudes which have power-law dependence on the soft energy, the logarithmic soft theorems are universal in the same sense as Weinberg's (leading) soft theorem. Hence their interpretation as conservation laws will lead us to the discovery of a new class of symmetries of scattering in four dimensions.

Building on our earlier results \cite{Choi:2024ygx}, we initiate a systematic analysis of the relationship between logarithmic soft photon theorem and asymptotic symmetries. Our goal in this paper is to show that the so-called classical logarithmic soft photon theorem  in (scalar) QED, which arises in the scattering of classical massive objects, is equivalent to the conservation law associated to  superphaserotations.\footnote{It is expected that a carefully taken classical limit of the full sub-leading factorization theorem in QED  will reproduce the classical log soft theorem. For recent progress in this direction see \cite{Manu:2020zxl, Alessio:2024onn}.} We will, moreover, prove that this conservation law is {\it exact to all orders in the electromagnetic coupling}. 

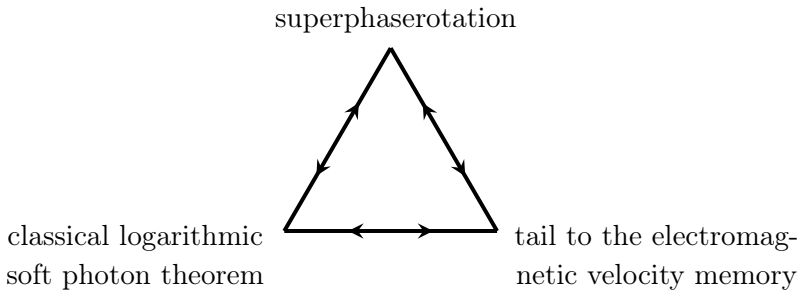
\begin{figure}[h!]
\begin{center}
  \tikzset{->-/.style={decoration={
    markings,
    mark=at position 0.7 with {\arrow{stealth}},
    mark=at position 0.3 with {\arrowreversed{stealth}}
    },
    postaction={decorate}}
  }
  \begin{tikzpicture}[scale=0.7]
    \coordinate (A) at (0,0);
    \coordinate (B) at (4,0);
    \coordinate (C) at (2,3.464);

    \node[text width=4cm,align=center] at (-2.8,-0.5) {classical logarithmic soft photon theorem};
    \node[text width=4cm,align=center] at (7,-0.5) {tail to the electromagnetic velocity memory};
    \node[text width=4cm,align=center] at (2.1,4) {superphaserotation};

    \draw [line width=1.5pt,->-] (A) -- (B);
    \draw [line width=1.5pt,->-] (B) -- (C);
    \draw [line width=1.5pt,->-] (C) -- (A);

  \end{tikzpicture}
  \end{center}
      \caption{Classical superphaserotation infrared triangle.}
      \label{fig:triangle}
\end{figure}

The infrared physics described by soft theorems and asymptotic symmetries has yet another incarnation in the form of memory effects. The classical logarithmic soft factor  \cite{Saha:2019tub} can also be interpreted as the so-called {\it tail} to the electromagnetic velocity memory~\cite{Saha:2019tub}. Here we generalize the formula for the tail memory to the case where matter is described by a massive scalar field. This completes the classical {\it infrared triangle} associated to superphaserotation symmetry which is summarised in Figure \ref{fig:triangle}.

In a companion paper \cite{CLP_Gravity}, we show that the universality of the logarithmic soft graviton theorem and the associated tail to the gravitational memory is due to the underlying superrotation symmetry of four dimensional asymptotically flat spacetimes; this yields the first infrared triangle in gravity where the long-range nature of gravitational interactions is taken into account. See also \cite{Agrawal:2023zea}.

We will now highlight the main results obtained in this work on infrared effects in scalar QED and place them in the context of known soft physics.
\paragraph{Power-law soft photon theorems.}

Scattering processes involving an external photon factorize when the momentum of the photon $k^\mu = \omega q^\mu$ is taken {\it soft}. 
The leading soft photon theorem is the statement that in the limit $\omega \to 0$ an $(N+1)$-particle amplitude factorizes into an $N$-particle amplitude without that soft photon times a soft factor that comes with a pole in the soft energy
\begin{equation}
\label{leadingsoft}
 \omega {\cal M}_{N+1}(\{p_{i}\};(\omega,q,\ell))=
   {S}_{-1} (\{p_{i}\};(q,\ell)){\cal M}_{N}(\{p_{i}\}),
   \end{equation}
The leading soft photon factor \cite{Weinberg:1965nx}
given by
\begin{equation}
\label{S-1}
    S_{-1}=e\sum_{i=1}^N  \frac{ Q_i p^\mu_i \varepsilon_\mu }{q\cdot p_i}
\end{equation}
is universal (theory-independent) as it only depends on the charges and momenta of the hard particles but not the details of the interactions. The leading soft theorem receives no corrections from loops.
There is a subleading soft photon theorem due to Low \cite{Low:1958sn} which can be expressed as
   \begin{equation}
   \label{subleadingsoft}
 (1+\omega \partial_\omega) {\cal M}_{N+1}(\{p_{i}\};(\omega,q,\ell))=
  {S}_{0} (\{p_{i}\};(q,\ell)){\cal M}_{N}(\{p_{i}\}),
   \end{equation}
with the subleading soft factor for minimally coupled scalar and spinor QED
\begin{equation}
\label{S0}
    S_{0}=-ie\sum_{i=1}^N  \frac{ Q_i q_\mu \varepsilon_\nu J_i^{\mu\nu}}{q\cdot p_i}.
\end{equation}
Unlike \eqref{S-1} the soft factor $S_0$ is not universal but gets corrected by (a short list of) effective operators with three-point interactions \cite{Elvang:2016qvq}. Moreover, loop corrections lead to the appearance of logarithms in the soft expansion which render the subleading tree-level soft theorem ambiguous. %

\paragraph{Logarithmic soft photon theorems.}

In scalar QED it was shown \cite{Sahoo:2018lxl,Saha:2019tub,Sahoo:2020ryf} that given the existence of a well-defined soft expansion to all orders in the loop expansion, the ratio between an $(N+1)$-point amplitude with a soft photon and the $N$-point amplitude without that photon is infrared-finite. Its soft expansion is given by
\begin{equation}\label{logsoftexp}
    \frac{{\cal M}_{N+1}(p_{1}, \dots, p_{N};(\omega,q,\ell))}{{\cal M}_{N}(p_{1}, \dots, p_{N})}=
    \sum_{n=-1}^{\infty} \omega^{n} (\ln\omega)^{n+1} {S}^{(\ln \omega)}_{n}+\dots
\end{equation}
The $n=-1$ term is the leading (Weinberg) soft factor ${S}^{(\ln \omega)}_{-1}\equiv {S}_{-1}$ while for $n\ge0$ the soft expansion differs from the tree-level one. It was shown in \cite{Sahoo:2018lxl,Sahoo:2020ryf} that not only the leading but also the subleading soft factor $S^{(\ln \omega)}_{0}\neq S_0$ is universal and uniquely fixed by the charges and the momenta of the scattering states.
The logarithmic soft factor may be split as \cite{Sahoo:2018lxl}
\begin{equation}
\label{Sln0}
    {S}^{(\ln \omega)}_{0}={S}^{(\ln \omega)}_{0,{\rm classical}}+\Delta{S}^{(\ln \omega)}_{0, {\rm quantum}}.
\end{equation}
The classical log soft photon factor is given by 
\begin{equation}
\label{Sln0sQEDclassical}
    {S}^{(\ln \omega)}_{0,{\rm classical}}=-i e^3\sum_{i=1}^N  \frac{\varepsilon_\mu  q_\nu Q_i}{ q\cdot p_i}\sum_{j\neq i, \eta_i \eta_j=1}\frac{Q_i Q_j}{4\pi} \frac{p_i^2p_j^2\left[p^\mu_i p^\nu_j-p^\mu_j p^\nu_i\right]}{\left[(p_i \cdot p_j)^2-p_i^2 p_j^2\right]^{3/2}}
\end{equation}
and can be derived from purely classical scattering. It has its origin in the late time electromagnetic radiation due to the acceleration of the charged particles via long-range electromagnetic interactions and is universal in that it is independent of the theory and the nature of external particles. 
In the quantum computation there is an additional contribution  
\begin{align}
\label{Sln0sQEDquantum}
\badat{2}
   \Delta {S}^{(\ln \omega)}_{0, {\rm quantum}}&=e^3\sum_{i=1}^N\frac{\varepsilon_\mu q_\nu Q_i}{ q\cdot p_i}\left(p^\mu_i\partial^\nu_{p_i}-p^\nu_i\partial^\mu_{p_i}\right)\\
    &\quad \sum_{j\neq i} \frac{Q_iQ_j}{8\pi^2} \frac{(p_i\cdot p_j)}{\sqrt{(p_i\cdot p_j)^2-p_i^2 p_j^2)}}\ln\left(\frac{p_i\cdot p_j+\sqrt{(p_i\cdot p_j)^2-p_i^2p_j^2}}{p_i\cdot p_j-\sqrt{(p_i\cdot p_j)^2-p_i^2p_j^2}}\right).
\eadat
\end{align}
This quantum log soft factor is also expected to be universal but has been computed directly from one-loop amplitudes in scalar QED and arises from the region where the loop momentum is large compared to the soft energy $\omega$ but small compared to the energies of the other particles.
In this work our goal is to derive a symmetry interpretation for the classical logarithmic soft photon factor \eqref{Sln0sQEDclassical}, while quantum infrared effects will be discussed elsewhere \cite{CLPquantumlog}.

\paragraph{Conservation laws for superphaserotations.}

To achieve a first-principles derivation of the classical logarithmic soft photon theorem from symmetries we need to construct finite superphaserotation charges and show that the associated conservation law implies the logarithmic soft theorem. We will work in the covariant phase space formalism where charges for asymptotic symmetries are extracted from the symplectic structure
\begin{equation}
\label{Omegapm}
  \Omega_{ i^{\pm}\cup {\cal I}^{\pm}}=\Omega^{\rm mat}_{i^{\pm}}+\Omega^{\rm rad}_{{\cal I}^{\pm}}.
\end{equation}
which has matter and radiative contributions from, respectively, time-like and null infinity. 
When one of the field variations is a superphaserotation, denoted here by $\delta_\ep$,
one obtains the variation of the charge from the symplectic structure
\begin{equation}
\Omega_{\,i^\pm\,\cup \,\cal I^\pm\,}(\delta,\delta_\ep)=\delta Q_{\pm}
.
\end{equation}
Moreover, in \cite{Lysov:2014csa,Campiglia:2016hvg}, the classical conservation law $Q_+=Q_-$
elevated at the level of the Ward identity $[Q,S]=0$ is shown to be equivalent to the tree-level subleading soft photon theorem.
In this paper, we revisit the derivation of the superphaserotation charges in four dimensions taking the infrared effects into account.

Due to the long-range nature of electromagnetic interactions asymptotic (charged) matter fields are not free but are `dressed' by a Coulombic phase \cite{Kulish:1970ut},
\begin{equation}
\label{phidressed}
    \phi=e^{i { \Phi}} \phi_{\rm free}.
\end{equation}
The dressing ${ \Phi}$ is logarithmically divergent at early and late times and is determined, via the equations of motion in conjunction with the asymptotic fall-offs of the electromagnetic field which have power-law `tails' at early and late times. These logs and tails lead to infrared divergences in the symplectic structure on the time-like and null boundaries which we regulate with an asymptotic cutoff $\Lambda^{-1}$.
From this we can then construct a regularized Noether charge
\begin{equation}
\label{NoetherQ}
  Q^{\Lambda}=
        \ln \Lambda^{-1}\, \left(Q^{(\ln)}_H+ Q^{(\ln)}_S \right) + \left(Q^{(0)}_H+ Q^{(0)}_S\right)+ \dots
\end{equation}
where the subscripts $H$ and $S$ refer to the `hard' and `soft' contributions to the total charge which originate from, respectively, the $i^\pm$ and the $\scri^\pm$ contributions to \eqref{Omegapm}. The first-principles derivation of this charge in scalar QED  is our main result. 

The finite charge 
\begin{equation}
\label{Q0}
    Q^{(0)}\equiv Q^{(0)}_H+ Q^{(0)}_S
\end{equation}
obeys a classical conservation law 
\begin{equation}
  Q^{(0)}_+=Q^{(0)}_-,
\end{equation}
which at the quantum level becomes the subleading tree-level soft theorem in scalar QED \eqref{subleadingsoft} with soft factor \eqref{S0}. Thus we rediscover the result of \cite{Campiglia:2016hvg} where infrared effects were ignored.
However, the charge \eqref{Q0} is ambiguous in the presence of long-range interactions.
On the other hand, the log charge
\begin{equation}
\label{Qlnintro}
 Q^{(\ln)}\equiv    Q^{(\ln)}_H + Q^{(\ln)}_S
\end{equation}
is unambiguous and we will show that its conservation law
\begin{equation}
\label{Qlnconservation}
 Q^{(\ln)}_+= Q^{(\ln)}_-
\end{equation}
reproduces precisely the classical logarithmic soft photon theorem \eqref{logsoftexp} with soft factor \eqref{Sln0sQEDclassical}.
Let us emphasize that in \cite{Campiglia:2019wxe} the conservation law \eqref{Qlnconservation} was reverse engineered from the log soft theorem which led the authors propose the charge \eqref{Qlnintro}. In \cite{Choi:2024ygx}, and in more detail here, we provide a first-principles covariant phase space derivation of this charge. We furthermore show in this work that the log charge \eqref{Qlnintro} is exact to all orders in the coupling $e$ and this echoes beautifully the one-loop exactness of the logarithmic soft photon theorem.

\section{Preliminaries}
\label{sec:Preliminaries}
The covariant phase space of the classical field theory is an infinite dimensional symplectic manifold with a symplectic structure $\Omega$ which is defined on \emph{any} Cauchy slice $\Sigma_{t}$ fixed at the inertial time $t$. In gauge theories and gravity, a convenient approach to obtain the gauge invariant symplectic structure is known as the approach of ``asymptotic quantisation''. In this approach we push forward the total symplectic structure (which is invariant under time evolution by the equations of motion) to the past and future boundaries, which is a union of time-like infinity $i^{\pm}$, and null infinity ${\cal I}^{\pm}$.  We review the approach to take these limits by starting with Minkowski spae-time in spherical coordinates
\begin{equation}
    ds^2=\eta_{\mu\nu}dx^\mu dx^\mu=-dt^2+d\vec x\cdot d\vec x,
\end{equation}
where $x^\mu=(t,\vec x)=(t,r \hat x)
$ with $r^2=\vec{x}\cdot \vec{x}$,
In this section, we briefly review the limiting procedure of taking $\Sigma_{t}\, \rightarrow\, i^{\pm} \cup {\cal I}^{\pm}$  that land us on \eqref{Omegapm}. 

\paragraph{Radiation.}
In order to compute the contribution of the radiative electromagnetic field at $\scri^+ \simeq S^2 \times \mathbb R$, we use the retarded Bondi coordinates $(u = t - r, r, \hat x)$ in which the flat metric takes the standard form, 
\begin{equation}
\label{ds2Bondi}
    ds^2=-du^2-2du dr+r^2\gamma_{AB} dx^A dx^B.
\end{equation}
Here $x^A=(z,\bar z)$ are the stereographic coordinates on the celestial sphere and  $\gamma_{AB}$ denotes the standard metric on $S^{2}$.\footnote{In Bondi coordinates, the flat metric has a non-trivial set of Christoffel symbols,
\begin{align}
    \Gamma^u_{AB} = r\gamma_{AB}
    ,\qquad
    \Gamma^r_{AB} = -r\gamma_{AB}
    ,\qquad
    \Gamma^A_{rB} = \frac1r \delta^A_B
    ,\qquad
    \Gamma^A_{BC} = {}^{(2)}\Gamma^A_{BC}
    ,
\end{align}
where ${}^{(2)}\Gamma^A_{BC}$ is the Christoffel symbol for the sphere, which has two non-vanishing components $\Gamma^z_{zz}=\frac{-2\zb}{1+z\zb}$ and $\Gamma^\zb_{\zb\zb}=\frac{-2z}{1+z\zb}$.}
A spacetime vector $x^{\mu}$ is parametrized in terms of basis vectors in the Bondi coordinates as,\footnote{
    We use the notation $\hat x$ for a spatial 3-vector restricted to the unit sphere.
    With a slight abuse of notation, we shall also use $\hat x$ to denote angular variables collectively: $d^2\hat x\equiv dzd\zb\gamma_{z\zb}$ is the volume element of $S^2$, and $f(\hat x)\equiv f(z,\zb)$ is a function on $S^2$.
}
\begin{equation}
    x^\mu=u(1,\vec0)+r(1,\hat x).
\end{equation}
The radiative contribution to the symplectic structure \eqref{Omegapm} is then computed by taking the $t$~=~constant time slice $\Sigma_t$, on which it is defined, to the future boundary  
\begin{equation}
\label{Omrad}   
  \Omega^{\rm rad}_{{\cal I}^{+}}=      \lim_{\Sigma_t\to \scri^+} \Omega_{t}^{\rm rad}.
\end{equation}
The corresponding contribution on the past boundary is computed in an analogous way in terms of the advanced Bondi coordinate $v = t + r$ which is held fixed as $t-r\to -\infty$.

\paragraph{Matter.}
Future and past time-like infinity are the asymptotic boundaries of any time-like trajectory. In other words, $i^{\pm}$ is reached in the limit $t\, \rightarrow\, \pm \infty$ while holding $\frac{r}{t}$ fixed.  $i^\pm$ can also be understood as the asymptotic limit of co-dimension one hyperbolic, or Euclidean AdS$_3$, slices which foliate the interior of the past and future light cones of any point inside Minkowski spacetime. 

The one-parameter family of coordinates inside the future light cone of the spacetime point $(u_0,\vec{0})$ %
is defined by
\begin{equation}
\label{taurho}
\tau = \sqrt{t^{2} - r^{2}} - u_{0}
,\qquad \tau \rho = r.
\end{equation}
It can be checked that any fixed $\tau$ hyperboloid intersects ${\cal I}^{+}\simeq S^2 \times \mathbb R$ at $u = u_{0}$. Due to translation symmetry of underlying spacetime, the choice of $u_{0}$ is arbitrary and following \cite{Campiglia:2015qka}, we will fix $u_{0} = 0$. Hence, $(\tau,\rho,x^A)$ coordinates cover the (interior) of the future light cone with tip at $(0,\vec{0})$. 
In these coordinates, the Minkowski spacetime is foliated by constant-$\tau$ space-like hyperboloid ${\cal H}_{\tau}$  
\begin{equation}
\label{ds2H3}
       ds^2=
        -d\tau^2
        + \tau^2 k_{\alpha\beta} dy^\alpha dy^\beta, 
\end{equation}
where $y^\alpha=(\rho,x^A)$ denote the coordinates on $\mathcal{H}_\tau$ which has an induced metric,\footnote{The non-vanishing Christoffel symbols for the metric~\eqref{ds2H3} are
\begin{align}
    \Gamma^\tau_{\alpha\beta} = \tau k_{\alpha\beta}
    ,\qquad
    \Gamma^\alpha_{\tau\beta} = \frac{1}{\tau} \delta^\alpha_\beta
    ,\qquad
    \Gamma^\alpha_{\beta\gamma} = {}^{(3)}\Gamma^\alpha_{\beta\gamma}
    ,
\end{align}
where ${}^{(3)}\Gamma^\alpha_{\beta\gamma}$ is the Christoffel symbol of the induced metric on the hyperboloid.}
\begin{equation}
    k_{\alpha\beta} dy^\alpha dy^\beta=
        \frac{d\rho^2}{1+\rho^2}
        + \rho^2\gamma_{AB}dx^Adx^B.
    \label{rhox}
\end{equation}

Let us define a vector
\begin{equation}\label{Y}
  \Y^\mu=(\sqrt{1+{\rho}^2},{\rho} \hat x)
\end{equation}
on Minkowski space which is parametrized by $y^\alpha=(\rho,x^A)$, normalized such that $\Y\cdot\Y=-1$. Any spacetime vector $x^{\mu}$ inside the light cone can then be written as, 
\begin{equation}
     {x}^\mu=\tau \Y^\mu.
\end{equation}
The `blow-up' of future time-like infinity $i^{+}$ is thus diffeomorphic to the Euclidean hyperboloid ${\cal H}$  approached via $\lim_{\tau\, \rightarrow\, \infty}\, {\cal H}_{\tau}$.  %
The contribution of the massive matter field to the symplectic structure \eqref{Omegapm} on the future boundary is then a limit of a one-parameter family of symplectic structures defined on ${\cal H}_{\tau}$,\footnote{We remind the reader that the Cauchy slice is diffeomorphic to ${\cal H}_{\tau}\, \cup\, (S^{2}\, \times\, {\bf R}^{-}$), where ${\cal R}^{-}$ is the negative half-line coordinatized by $u\, \in\, (-\infty, 0]$. As massive field decays exponentially fast as we approach null infinity, the matter symplectic structure only has support on ${\cal H}_{\tau}$ \cite{Campiglia:2015qka}.}
\begin{equation}
\label{Ommat}
 \Omega^{\rm mat}_{i^{+}}=\lim_{{\cal H}_\tau\to i^+}\Omega_{\tau}^\text{mat}.
\end{equation}
Similarly, the matter symplectic structure on the past boundary can be computed and has support on the blow up of $i^{-}$, where $i^{-}$ is coordinatized by $(\rho, x^{A})$ and is related to the advanced Bondi coordinates via the map,
\begin{align}
    \tau\, =\, -\sqrt{v^{2} + 2 v r},\qquad
    \rho\, =\, \frac{r}{|\tau|}
    .
\end{align}

The sum of the radiative and matter symplectic structures in \eqref{Omrad} and \eqref{Ommat} is then the definition of the complete symplectic structure on the covariant phase space and is parametrized in terms of asymptotic data at $i^{+}\, \cup\, {\cal I}^{+}$. 

The intersection of $i^{+}$ with ${\cal I}^{+}$ is the co-dimension one boundary of ${\cal I}^{+}$ which is the celestial sphere $S^{2}$ as $u\, \rightarrow\, \infty$. This boundary of ${\cal I}^{+}$ is denoted in the literature as ${\cal I}^{+}_{+}$.   
The intersection ${\cal I}^{+}_{+} := {\cal I}^{+} \cap\, i^{+}$ will allow us to geometrically relate the asymptotic (late time) expansions of the fields at null infinity with the asymptotic field at $i^{+}$. All of the above statements continue to hold for the past boundary, $i^{-}\, \cup\, {\cal I}^{-}$.

In the following we will derive explicit expressions for \eqref{Omrad} and \eqref{Ommat} for scalar QED.
In order to proceed, we will require the following data :  $1)$ the asymptotic phase space of scalar QED  with the suitable boundary conditions for the charged scalar field and the Maxwell field;  $2)$ the asymptotic symmetry action on fields at $i^{\pm}\cup {\cal I}^{\pm}$.
For the benefit of the reader, we discuss each of these ingredients separately in detail.

\section{Asymptotic data in scalar QED}
\label{sec:Asymptotics}
To compute the matter and radiative contributions to the symplectic structure in scalar QED
we need to establish the asymptotic phase space.
The field content is a massive complex scalar $\phi$ coupled minimally to a gauge field $\calA_\mu$ with action
\begin{equation}
    S
    =
        -\int d^4x\left\{
            \left[(\nabla_\mu - ie\calA_\mu)\phi\right]^*
            (\nabla^\mu-ie\calA^\mu)\phi
            + m^2\phi^*\phi
            + \frac14 \calF_{\mu\nu}\calF^{\mu\nu}
        \right\}.
\end{equation}
The equations of motion are
\begin{equation}
      (\nabla_\mu - ie \calA_\mu)(\nabla^\mu - ie \calA^\mu) \phi-m^2\phi=0,
   \qquad \nabla^\nu \mathcal F_{\mu\nu}=j_\mu,
\end{equation}
with %
field strength $\mathcal F_{\mu\nu}=\partial_\mu \mathcal A_\nu-\partial_\nu \mathcal A_\mu$, and Noether current for the matter field
\begin{equation}
 j_\mu=i e\phi (\nabla_\mu + ie \calA_\mu) \phi^* +{\rm c.c.}\,.
\end{equation}
We will work in Lorenz (harmonic) gauge $\nabla^\mu \calA_\mu=0$. 
In turn we will discuss the asymptotic fall-offs for the charged scalar matter and the gauge field. We will focus on the asymptotics on $i^+ \cup \scri^+$, but a similar analysis can be repeated for $i^- \cup \scri^-$.

\subsection{Matter}

The late time asymptotic behaviour of a massive scalar field  was analysed in \cite{Campiglia:2015qka,Campiglia:2019wxe}. Starting from a free matter field the interaction with the photon field leads to a logarithmic dressing of the former and a tail at late times in the latter. In the following we will derive the late-time expressions of the matter and photon fields to all orders in the coupling $e$ thus extending the perturbative results of \cite{Campiglia:2015qka,Campiglia:2019wxe}. 
We use hyperbolic coordinates $(\tau,y^\alpha)=(\tau,\rho, x^A)$. 
The Lorenz gauge condition takes the form
\begin{equation}
\label{Lorenzi+}
\badat{2}
    0&=%
    - \left(\p_\tau 
		+ \frac3{\tau} \right) \calA_\tau
		+ \frac1{\tau^2}{\cal D}^\alpha \calA_\alpha,
\eadat
\end{equation}
where ${\cal D}_{\alpha}$ is the covariant derivative in the three-dimensional hyperbolic space. The equation of motion for the scalar field is%
\begin{equation}
\label{sQEDphieom}
\badat{2}
	0=
        \left[- \partial_\tau^2 
        - \frac3\tau \partial_\tau 
        + \frac1{\tau^2}{\cal D}^2-m^2
        + 2ie\calA_\tau \left(\partial_\tau -\frac{ie}{2}\calA_\tau\right)
        - \frac{2ie}{\tau^2} \calA_\alpha\left({\cal D}^\alpha-\frac{ie}{2} \calA^\alpha\right)\right]\phi
\eadat
\end{equation}
where ${\cal D}^2\equiv{\cal D}_\alpha{\cal D}^\alpha$. While there is no radiation at $i^\pm$ the %
matter current $j_\mu$ sources a `Coulombic' gauge field  via Maxwell's equations %
which in hyperbolic coordinates take the form %
\begin{equation}
\label{Maxwell}
   \badat{2}
j_\tau&=\left[\partial_\tau^2 +\frac{5}{\tau}\partial_\tau -\frac{1}{\tau^2} (\calD^2 -3)\right]\calA_\tau,\\
j_\alpha&=\left[\partial_\tau^2 +\frac{1}{\tau}\partial_\tau  -\frac{1}{\tau^2} (\calD^2 +2)\right]\calA_\alpha+\frac{2}{\tau} \calD_\alpha A_\tau.\\
   \eadat
\end{equation}

The asymptotics of a {\it free} complex massive scalar field in Minkowski space,
\begin{equation}
\label{freescalar}
    \phi_\text{\rm free}(x)=\int \frac{d^3\vec{p}}{(2\pi)^3 2E_p} \left[b(\vec{p}) e^{ip\cdot x}+d^*(\vec{p})e^{-ip\cdot x}\right]
\end{equation}
with $E_p=\sqrt{|\vec{p}|^2+m^2}$ and $p\cdot x=-E_pt+\vec{p}\cdot \vec{x}$, is described along time-like geodesics and its late-time ($\tau \to \infty$) behavior can be extracted via a saddle point approximation with critical point $\vec{p}=m\rho\hat{x}$%
.
After absorbing various %
phases into the `free data' given by $b$ and $d^*$, the late-time asymptotics of the free scalar field can be expressed as
\begin{equation}
    \label{phi0}
    \phi_\text{\rm free}(\tau, y) \stackrel{\tau\to\infty}{=} 
    \frac{\sqrt m}{2(2\pi\tau)^{\frac{3}{2}}}
    \left[ b(y) e^{- i m \tau} +  d^{*}(y) e^{im\tau}  \right].
\end{equation}
where $b(y)\equiv b(m\rho\hat x)$.
Via Maxwell's equations \eqref{Maxwell} the free matter current  
\begin{equation}
    j^{\rm free}_\mu=ie\phi_\text{\rm free}\partial_\mu \phi_\text{\rm free}^*+c.c.
\end{equation}
sources a Coulombic potential. In particular, its $\tau$-component is of order $O(1/\tau^3)$ and thus %
sources a $1/\tau$ late time `tail' in the gauge potential \footnote{This is familiar from the Coulomb potential sourced by a massive point particle which at late times for asymptotic constant particle velocity
becomes $\frac{Q}{4\pi \tau}$. See \cite{Laddha:2018myi,Laddha:2018vbn,Sahoo:2018lxl} for recent classical results on the late-time logarithmic behavior of fields due to long-range interactions.
}
\begin{equation}
\label{Atauleading}
    \calA_\tau(\tau,y)\stackrel{\tau \to\infty}{=} \frac{1}{\tau} \overset1A_\tau(y)+\dots
\end{equation}
where $\mathcal D^2 \overset1A_{\tau}=-\overset{3}{j_\tau}$.\footnote{Our notation differs from that in \cite{Campiglia:2015qka,Campiglia:2019wxe}. Here a superscript $n$ denotes the coefficient of the $\frac{1}{\tau^n}$ term.}
The Lorenz gauge condition \eqref{Lorenzi+} then implies
\begin{equation}
\label{Aalphaleading}
    \calA_\alpha(\tau,y)\stackrel{\tau \to\infty}{=}  \overset0A_\alpha(y)+\dots
\end{equation}
where $\mathcal D^\alpha \overset0A_\alpha=2\overset1A_{\tau}$.\footnote{Compatibility with Maxwell's equations which imply $\left(\mathcal D^2+2\right) \overset0 A_\alpha= 2\mathcal D_\alpha \overset 1 A_\tau$ requires ${\cal D}^\beta\mathcal D_{[\alpha}\overset0A_{\beta]}=0$.}
Both Coulombic modes arise from the interactions so that $\overset1 A_\tau=O(e)$ and $\overset0 A_\alpha=O(e)$.
Plugging this asymptotic behavior into the scalar equation of motion \eqref{sQEDphieom} reveals that the free field gets corrected by a term $i e \ln \tau \overset1 A_\tau \phi_\text{\rm free}$ that dominates at late times. Indeed this follows from the leading correction to the free equation of motion $(\nabla^2-m^2+2ie\overset1A_\tau \tau^{-1}\partial_\tau+\dots)\phi=0$ at late times where the matter field is no longer free but gets `dressed' by
a logarithmic phase %
\begin{equation}
\label{phifreedressed}
      \phi(\tau, y)
    \stackrel{\tau \to\infty}{=} e^{i e \ln \tau \overset1 A_\tau}\phi_\text{\rm free}(\tau,y) +\dots \,.
\end{equation}
While our discussion so far implies only corrections quadratic in the coupling, we will show in Appendix~\ref{app:sQEDallorder} that the dressing \eqref{phifreedressed} is actually exact to all orders in $e$. In fact we will find that the infrared corrected charge we construct from the matter symplectic structure is exact in the coupling with no corrections beyond cubic order in $e$. 
For ease of presentation we will in the following restrict our discussion to those orders in the coupling that eventually contribute to the charge and relegate the details of the all-order derivation to Appendix~\ref{app:sQEDallorder}.

Expanding the complex scalar field up to second order in the coupling and keeping only the most relevant terms in the large-$\tau$ expansion,
\begin{equation}
\label{phiati+}
    \phi(\tau, y)
    \stackrel{\tau \to\infty}{=} 
	    \frac{e^{-im\tau}}{\tau^{3/2}}
	    \left(b_0(y)+
	    	\overset \ln b{}_{0}(y)\ln\tau
	    \right)
     + \frac{e^{-im\tau}}{\tau^{5/2}}
	    \left(b_1(y)+
	    	\overset \ln b{}_{1}(y)\ln\tau
	    \right)
    +... +O(e^{3}),
\end{equation}
we can derive a series of algebraic relations between its coefficients and the Coulomb field by solving the equation of motion \eqref{sQEDphieom} at each order in $\tau$. 
The result is
\begin{equation}
    \label{QED_scalar_expansion}
    \badat{2}
    & \qquad\qquad \qquad b_0,\quad b_1= \frac1{2im}
        \left(
             {\cal D}^2+\frac34
        \right)b_0+O(e^2),\\
    &\overset\ln b_0=ie\overset1A_\tau b_0
    +O(e^3), \quad \overset\ln b_1=
        \frac{e}{2m}
        \left(
            {\cal D}^2+\frac34
        \right) (\overset1A_\tau b_0)+O(e^3),
    \eadat
\end{equation}
and similar relations for the negative frequency modes $\overset\ln d{}^*_0$, $\overset\ln d{}^*_1$, $d^*_1$ which we have omitted in \eqref{phiati+} but which can easily be reinstated upon complex conjugation and replacing $m$ by $-m$. 
This procedure allows us to express the coefficients $\overset\ln b_n$ and $b_{n+1}$ ($\overset\ln d{}_n^*$ and $d^*_{n+1}$) for all $n\geq 0$ in terms of the lowest order ones $b_0 \equiv \sqrt{\frac{m}{4(2\pi)^3}}\,b$ ($d^*_0 \equiv  \sqrt{\frac{m}{4(2\pi)^3}}\,d^* $) %
which are precisely the free asymptotic data; in Appendix~\ref{app:sQEDallorder} we will complete \eqref{QED_scalar_expansion} to expressions that are exact in $e$. 

The late-time behavior of the dressed matter field at $i^+$  determines the asymptotic form of the matter current 
\begin{equation} 
\label{jtaualphai+}
\badat{2}
        j_\tau(\tau,y)
    &\stackrel{\tau \to\infty}{=}\frac{1}{\tau^3} \overset{3}{j_\tau}(y)+\frac{\ln \tau}{\tau^4} \overset{4,\ln}{j_\tau}(y)+\frac{1}{\tau^4} \overset4{j_\tau}
    +\dots\,,\\
    j_\alpha(\tau,y)
    &\stackrel{\tau \to\infty}{=}\frac{\ln \tau}{\tau^3} \overset{3,\ln}{j_\alpha}(y)+\frac{1}{\tau^3} \overset3{j_\alpha}+\dots\, .
\eadat
\end{equation}
Notice that while $j_\tau$ at leading order is given by the free matter current, $j_\alpha$ receives a logarithmic correction that dominates over the free expression.\footnote{Naively, also $j_\tau$ receives a leading logarithmic correction %
but the coefficient vanishes upon using \eqref{QED_scalar_expansion}. Moreover, to leading order in $1/\tau$, potential higher powers of $\ln \tau$ are absent as we will show in Appendix \ref{app:sQEDallorder}.}
The explicit expressions for the coefficients are
\begin{equation}
\badat2
    \overset3j_\tau
    &=
        -2em(b_0^*b_0-d_0^*d_0)
    ,\\
    \overset{4,\ln}{j_\tau}
    &=
        -2e^2 \calD^\alpha\Big[(\p_\alpha\overset1A_\tau)(b_0^*b_0+d_0^*d_0)\Big]
    ,\\
    \overset{3,\ln}{j_\alpha}
    &=
        2e^2(b_0^*b_0+d_0^*d_0) \p_\alpha \overset1A_\tau
    ,\\
    \overset3j_\alpha
    &=
        - ie\left(
            b_0^*\p_\alpha b_0
            - d_0^* \p_\alpha d_0
            - b_0 \p_\alpha b_0^*
            + d_0\p_\alpha d_0^*
        \right)
        - 2e^2 \overset0A_\alpha (b_0^*b_0+d_0^*d_0)
    .
\eadat
\end{equation}
While from the above discussion the asymptotic behavior of the current holds at $O(e^3)$, we will show in Appendix~\ref{app:sQEDallorder} that \eqref{jtaualphai+} in fact holds to all orders. In turn these logarithmic corrections turn on similar terms in the components \eqref{Atauleading}-\eqref{Aalphaleading} of the Coulomb field at subleading order.

We have focused here on the $\tau \to +\infty$ limit to reach $i^+$, but an analogous analysis can be performed for $\tau \to -\infty$ limit to reach $i^-$.
Finally, note that the asymptotic solution to \eqref{sQEDphieom} can be recast as a free scalar field dressed by a Wilson line for the Coulombic gauge potential 
\begin{equation}
    \phi(\tau,y)\stackrel{\tau \to \pm \infty}{=}e^{ie\int^\tau A_\tau(\tau',y)d\tau'} \phi_\text{\rm free}(\tau,y).
\end{equation}
In the quantum theory this corresponds to the Faddeev-Kulish asymptotics of charged quantum fields \cite{Kulish:1970ut}. Here this dressing was derived from a purely classical analysis. We emphasize again, that the matter dressing is not a choice, it is forced on us by the long-range nature of the electromagnetic interactions and thus must be accounted for in any asymptotic symmetry analysis. Moreover,   as demonstrated in Appendix \ref{app:sQEDallorder}, this dressing is exact in the coupling $e$. We will see that this has profound implications for the relation between asymptotic symmetries and the logarithmic soft photon theorem.

\subsection{Radiation}

At future (past) null infinity $\scri^+$ ($\scri^-$) we use retarded (advanced) Bondi coordinates $(u\,(v), r, x^A)$ and study the asymptotics of the radiative photon field at early/late times $u\to \pm \infty$ ($v\to \pm\infty$). We will focus on the future boundary here and only highlight the most pertinent aspects for our purpose. We refer to \cite{Campiglia:2015qka,Campiglia:2019wxe} for more details.

In retarded Bondi coordinates the Lorenz gauge condition is
\begin{equation}
\label{Lorenzscri}
    \partial_r\left(r^2\mathcal A_u\right) -\partial_r\left(r^2\mathcal A_r\right)+r^2\partial_u\mathcal A_r- D^C \mathcal A_C=0,
\end{equation}
and Maxwell's equations, $\nabla^2 \mathcal A_\mu=-j_\mu$,  take the form 
\begin{equation}
\label{Maxwellscri}
\badat{3}
j_u&=\Big[-\partial_r^2+2\partial_r \partial_u+\frac{2}{r} (\partial_u-\partial_r) -\frac{1}{r^2}D^2\Big] \mathcal A_u,\\
j_r&=\Big[-\partial_r^2+2\partial_r \partial_u+\frac{4}{r}(\partial_u-\partial_r) -\frac{1}{r^2}(D^2+2)\Big] \mathcal A_r+\Big[\frac{2}{r}\partial_r+\frac{2}{r^2}\Big]\mathcal A_u\\
j_A&=\Big[-\partial_r^2+2\partial_r \partial_u -\frac{1}{r^2}(D^2-1)\Big] \mathcal A_A+\frac{2}{r}D_A(\mathcal A_u-\mathcal A_r).
\eadat
\end{equation}

The asymptotics of the {\it free} Maxwell field in Minkowski space, 
\begin{equation}
    \calA^{\rm free}_\mu(x)=\int \frac{d^3k}{(2\pi)^32\omega_k}\left[a_{\mu}(\vec k)e^{ik\cdot x}+{a}_{\mu}(\vec k)^\dagger e^{-ik\cdot x}\right]
\end{equation}
with $\omega_k=k^0=|\vec k|$, $k\cdot x= -\omega_k t+\vec k \cdot \vec x$ and $a_{\mu}(\vec k)=\sum_{\lambda=\pm} \varepsilon^{\lambda*}_\mu a_\lambda(\vec k)$, can be obtained from a saddle point analysis with saddle point $\vec k=\omega_k\hat{x}$. The $r\to \infty$ limit (at fixed $u$) is
\begin{equation}
    \calA^{\rm free}_\mu(x)=-\frac{i}{8\pi^2r}\int_0^\infty d\omega_k \left[a_\mu(\omega_k,x^A)e^{-i\omega_k u}-a_\mu(\omega_k,x^A)^\dagger e^{i\omega_k u}\right]+O(1/r^2).
\end{equation}
The free data is given by the two independent polarizations of the photon.
After mapping the Cartesian components to Bondi coordinates, we find the asymptotic behavior 
\begin{equation}
    \calA^{\rm free}_u=O(r^{-1}),\quad \calA^{\rm free}_r=0,\quad \calA^{\rm free}_C=O(r^0).
\end{equation}
What about interactions? It turns out that long-range interactions modify the large-$r$ behavior at {\it overleading} order by a term logarithmic in $r$ \cite{Campiglia:2019wxe}. However, the field equations at null infinity imply that this more leading term is $u$-independent and so we may discard it for the purpose of our discussion.
Thus we can focus on deriving the effect of long-range interactions on the $u\to \pm \infty$ behavior. 

We can infer the late-time fall-offs at the future boundary of $\scri^+$ 
from the analysis of the matter field asymptotics at $i^+$.
Following the analysis of \cite{Campiglia:2019wxe}, we use the retarded Green's function for Maxwell's equations in Lorenz gauge sourced by the matter current to write,
\begin{equation}
\label{Aretarded} 
\mathcal A_\mu(x)=\frac{1}{2\pi}\int d^4x' \Theta(t-t')\delta\Big((x-x')^2\Big)j_\mu(x'),
\end{equation}
and evaluate it at $\scri^+_+$ by first taking $r\to \infty$ at fixed $u$ and then letting $u\to \infty$ where the photon field gets sourced by the late-time asymptotics of the matter current at $i^+$.
To that end we express the spacetime point $x^\mu$ in Bondi coordinates 
while we write ${x'}^\mu$ in hyperbolic coordinates
using the expressions in section \ref{sec:Preliminaries}.
At large $r$ and fixed $u$ we have
\begin{equation}
    -(x-x')^2=2r(u+\tau' q\cdot \Y')+O(r^0)
\end{equation}
where $q^\mu=(1,\hat x)$. Using $d^4x'=d^3y' d\tau' {\tau'}^3$ we have to evaluate
\begin{equation}
\label{Aretardedfin} 
\mathcal A_\mu(x)=\frac{1}{4\pi r}\int d^3y' d\tau'{\tau'}^3 \delta(u+\tau' q\cdot \Y')j_\mu(x').
\end{equation}
The Cartesian components of the current are related to the hyperbolic ones by
\begin{equation}
    j_\mu(x)=-\Y_\mu j_\tau+\frac{1}{\tau }\mathcal D^\alpha \Y_\mu j_\alpha.
\end{equation}
Inserting the large-$\tau$ behavior of the matter current \eqref{jtaualphai+} at $i^+$ we can thus extract the large-$u$ behavior of the Cartesian components of the photon field at $\scri^+$ as follows\footnote{We will show in Appendix \ref{app:sQEDallorder} that there are no higher powers of logarithms in the asymptotics of the photon field \eqref{Amuru}.}
\begin{equation}
\label{Amuru}
    \mathcal A_\mu(x)=\frac{1}{r}\Big[\overset0{A_\mu}(x^A)+\frac{\ln u}{u} \overset{1,\ln}{ A_\mu}(x^A)+\frac{1}{u}\overset1{A_\mu}(x^A)+\dots\Big].
\end{equation}
Using current conservation $\nabla^\mu j_\mu=0$ one finds (see Appendix \ref{app:ACufalloff}) that the coefficient of the $\ln u/u$ term vanishes, while the constant term is sourced by the free matter current
 \begin{equation}
     \overset0{A_\mu}(x^A)=\frac{1}{4\pi}\int d^3y (q\cdot \Y)^{-1} \Y_\mu \overset3{j_\tau},
 \end{equation}
 and the $1/u$ term is determined by the leading correction to the free matter current due to long-range interactions
\begin{equation}
    \overset1{A_\mu}(x^A)=\frac{1}{4\pi}\int d^3y (q\cdot \Y)^{-1}
    q^\nu J^\alpha_{\mu\nu}(y)\overset{3,\ln}{j_\alpha}(y),
\end{equation}
where $J^\alpha_{\mu\nu}\equiv \Y_\mu \mathcal D^\alpha \Y_\nu-\Y_\nu \mathcal D^\alpha \Y_\mu$. The relevant Bondi components are $A_C=\partial_C q^\mu A_\mu$. 
Let us summarize these results for the radiation field at null infinity.

The free data of the photon field at $\scri^+$ is given by 
\begin{equation}
\label{calAClarger}
    \calA_C(u,r,x^A)\stackrel{r\to \infty}{=} A_C(u,x^A)+\dots\,,
\end{equation}
while the Lorenz condition \eqref{Lorenzscri} implies
\begin{equation}
\label{calAurlarger}
\mathcal A_u(u,r,x^A)\stackrel{r\to \infty}{=}\frac{1}{r} A_u(u,x^A)+\dots,\quad \mathcal A_r(u,r,x^A)\stackrel{r\to \infty}{=}\frac{1}{r^2}A_r(u,x^A) +\dots.
\end{equation}
At early/late times $u\to \pm\infty$ the free data behaves as
\begin{align}
\label{ACufalloff}
        A_C(u, x^A)
        &\overset{u\to\pm\infty}=
            A_C^{(0),\pm}(x^A)
            + \frac{1}{|u|} A_C^{(1),\pm}(x^A)
            + O\Big(\frac{1}{ u^{1+\#}}\Big)
\end{align}
where  $\#$ is ({\it a priori}) any positive number\footnote{Restrictions will be placed on this number when matching asymptotic symmetries to soft theorems - we will get back to this in section \ref{sec:sQEDcharges}.}. The $1/\tau$ tail of the Coulombic mode $A_\tau$ thus has a null counterpart in the $1/u$ behavior of $A_C$. While the constant term is sourced by the free matter current, the $1/u$ term arises due to the long-range nature of the electromagnetic interaction at early/late times $u\to \pm\infty$,
\begin{equation}\label{AC0+AC1intermsofj}
    A_C^{(0),\pm}=\frac{1}{4\pi} \int_{i^\pm} d^3 y \frac{\partial_C q\cdot \Y}{q\cdot \Y} \overset{3}{j_\tau}(y),\qquad  A_C^{(1),\pm}=\frac{1}{4\pi} \int_{i^\pm} d^3 y \frac{\partial_C q^\mu q^\nu J^\alpha_{\mu\nu}}{q\cdot \Y} \overset{3,\ln}{j_\alpha}(y).
\end{equation}
We can recognize the right hand side of $A_{C}^{(0),\pm}$ with the velocity memory for EM field sourced by a charged massive field. This can be deduced simply via the following identification. For the massive scalar field, the asymptotic current and momentum densities at $i^{+}$ are
\begin{align}
J^{\mu}(y)\, :=\, P^{\mu}\, \overset{3}{j}_{\tau}(y),\qquad
P^{\mu}(y)\, :=\, {\cal Y}^{\mu} \overset3T_{\tau\tau}(y),
\end{align}
respectively, where $\overset3T_{\tau\tau}(y)$ is the energy density of the scalar field at $i^{+}$ and is defined as follows\footnote{It can be readily verified that the expression \eqref{t3tt} is the energy density because the Poisson bracket $i\{b(\vec p),b^*(\vec p')\}=(2\pi)^3(2E_p)\delta^3(\vec p-\vec p')$ corresponds to $i\{b(y),b^*(y)\}=(2\pi)^3(2m^{-2})\delta^3(y-y')$, so after quantization the action of $\overset3T_{\tau\tau}$ on a massive single-particle state is $T_{\tau\tau}\ket{\vec p}=m\ket{\vec p}$.} \cite{Campiglia:2015kxa}
\begin{align}
\overset3T_{\tau\tau}(y) = \frac{m^3}{2(2\pi)^3} b^*(y) b(y).
\label{t3tt}
\end{align}
In terms of $J^{\mu}(y)$ we can write $A_{C}^{(0),+}$ as
\begin{align}
A_{C}^{(0),+}\, =\, \frac{1}{4\pi}\, \int_{i^+} d^{3} y \frac{\partial_{C} q \cdot J(y)}{q \cdot P(y)}
,
\end{align}
which generalizes the formula for the radiative gauge field sourced by point particles as $u\to+\infty$,
\begin{align}
A_{C}^{(0),+}\, =\, \sum_{i}\, \frac{(\partial_{C}q) \cdot (Q_{i} p_{i})}{q \cdot p_{i}}
,
\end{align}
where $Q_{i}$ is the charge of the $i$-th particle with asymptotic momentum $p_{i}^{\mu}$.\footnote{In a slight abuse of notation, we identify $Q p^{\mu}$ as current instead of $\frac{1}{m}Q p^{\mu}$.}
This generalizes the formula for electromagnetic velocity memory \begin{equation}
    \Delta A_C^{(0)}
    =
        A_C^{(0),+}-A_C^{(0),-}
\end{equation}
from particles to massive fields.
We similarly expect
\begin{equation}
    \Delta A_C^{(1)}
    =
        A_C^{(1),+}-A_C^{(1),-}
\end{equation}
to be the field-theoretical counterpart of the tail to the velocity memory; a detailed study is left for future work.

\section{Superphaserotation symmetry}
\label{sec:superphaserotations}
Now that we have identified the asymptotics of the fields at boundary arising from the long-range nature of the interactions, our next goal is to find a symmetry transformation acting on the fields that is consistent with these logs and tails.
We start by identifying a suitable symmetry parameter $\epsilon$ in the transformation of the gauge field $\calA_\mu$ on $\scri^+$ and of the complex massive scalar field $\phi$ on $i^+$,
\begin{equation}\label{deltaAphi}
\delta {\cal A}_\mu=\partial_\mu \epsilon^{\cal I^\pm},\quad \delta \phi=i e \epsilon^{i^\pm} \phi.
\end{equation}
Such transformations are called `large' when the symmetry parameter decays slow enough at the boundary to yield non-vanishing Noether charges such that they are not gauge redundancies but symmetries.
We will hitherto refer to the transformation \eqref{deltaAphi} as superphaserotation. 
In Lorenz gauge the superphaserotation parameter satisfies 
\begin{equation}
\label{SPRinLorenz}
    \nabla^2 \epsilon^{\cal I^\pm}= \nabla^2 \epsilon^{i^\pm}=0.
\end{equation} 
We will now identify, consistent with long-range interactions, superphaserotations with parameters $\epsilon^{i^+}$ on the time-like and $\epsilon^{\scri^+}$ on the null boundary and how to smoothly connect them. 
The leading soft photon theorem can be recast as a Ward identity for `leading' superphaserotations whose symmetry parameters at $i^\pm\cup \scri^\pm$ depend only on the angles $x^A$ of the $S^2$. The `subleading' superphaserotations associated to the subleading tree-level soft photon theorem are divergent: as $O(\tau)$ on $i^\pm$ and as $O(r)$ on $\scri^\pm$. We expect a similarly divergent behavior for the symmetry parameters associated with the logarithmic soft photon theorem which we now discuss. Again we will focus on the future boundary, but an analogous discussion holds for the past boundary.

\paragraph{Superphaserotation on $\scri^+$.}

Our ansatz for the large-$r$ limit of the superphaserotation parameter on $\scri^+$ is \cite{Campiglia:2016hvg}
\begin{equation}
\label{epsscri+}
   \epsilon^{\scri^+}(u,r,\hat x)
    \stackrel{r\to \infty}{=}
        \left[r
        + \frac{u}{2}\left(D^2+2\right)\right]\epsilon(\hat x) + O\left(\tfrac{\ln r}{r}\right)
    ,
\end{equation}
where $D^2$ is the Laplacian on the $S^2$. The form of the term linear in $u$ is required to solve \eqref{SPRinLorenz}. Its action on the Maxwell field \eqref{calAClarger} is
\begin{equation}
    \label{SPRAC}
   \delta \calA_C(u,r,\hat x) = \partial_C\left\{\left[r 
   + \frac{u}{2}(D^2+2)\right]\epsilon(\hat x)\right\} +\dots
\end{equation}
We may think of this as shifting the `background field' by terms that diverge as $O(r)$ and $O(u)$ at large distance and early/late retarded time,
\begin{equation}
    \label{SPRandShift}
   \calA_C(u,r,\hat x) \mapsto \partial_C\left[r \alpha(\hat x)  + u\beta(\hat x) \right]+A_C(u,\hat x)+\dots\,.
\end{equation}
The Goldstone modes $\alpha$ and $\beta$ transform homogeneously,
\begin{align}
\label{Goldstonedelta}
    \delta\alpha = \epsilon
    ,\qquad
    \delta\beta = \left(1+\tfrac{D^2}{2}\right)\epsilon,
\end{align}
and cannot be switched off. %

\paragraph{Superphaserotation on $i^+$.}

Our ansatz for the large-$\tau$ behavior of the superphaserotation parameter on $i^+$ is
\begin{equation}
\label{epsi+}
    \epsilon^{i^+}(\tau,y)
    \stackrel{\tau \to \infty}{=}
        \tau\bar\ep(y)
        + O\left(\tfrac{\ln\tau}{\tau}\right)
    .
\end{equation}
In Lorenz gauge \eqref{SPRinLorenz} is satisfied as
\begin{equation}
    ({\cal D}^2- 3)\bar\ep(y)=0.
\end{equation}
Its action on complex matter field is
\begin{equation}
\label{deltaphi}
    \delta \phi(\tau,y)=i e \,\tau \bar \ep(y) \phi(\tau,y)+\dots\,,
\end{equation}
corresponding to a rotation by an $O(\tau)$ divergent phase at early/late times.

\paragraph{Superphaserotation on $i^+\cup \scri^+$.}
The final step is to identify a superphaserotation parameter that smoothly interpolates across the union of the future time-like and null boundaries $i^+\cup \scri^+$. First note that we may parameterize $\bar\ep(y)$ on $i^+$ in terms of a function $\epsilon(\hat x)$ on the $S^2$ \cite{Campiglia:2015lxa}
\begin{equation}
    \bar\ep(y)=
        \int_{S^2} d^2\hat x\,
        G(y; \hat{x})\, \ep(\hat{x}),
\end{equation}
where $G(y; \hat{x})$ is the bulk-to-boundary Green's function which satisfies 
\begin{equation}
    ({\cal D}^2 - 3 ) G(\rho,\hat x;\hat x')
    =0.
\end{equation}
In the large-$\rho$ limit its asymptotic behavior is
\begin{equation}
    \lim_{\rho\to\infty} \rho^{-1}
    G(\rho,\hat x;\hat{x}') = \delta^{2}(\hat x-\hat x').
\end{equation}
Then using \eqref{taurho} we find that at the boundary of $i^+$ the superphaserotation parameter is
\begin{equation}
     \epsilon^{i^+}(\tau,\rho,\hat x)
    \overset{\rho\to \infty,\tau\to \infty}{=} r \epsilon(\hat x)+\dots
\end{equation}
Upon choosing the same function $\epsilon(\hat x)$ on the $S^2$ for the superphaserotation parameters on $i^+$ and $\scri^+$ we have that at leading order
\begin{equation}
    \epsilon^{i^+}=\epsilon^{\scri^+}.
\end{equation}
We have thus identified a superphaserotation parameter that smoothly extends across the future boundary, $i^+\cup \scri^+$, with the one at $i^+$ being `sourced' by the one at $\scri^+$.

\section{Symplectic structure for scalar QED}
Given the fall-offs of the matter and radiative gauge fields at time-like and null infinity, we are now in the position to compute the total symplectic structure of scalar QED, 
\begin{equation}
    \Omega_{i^+\cup \scri^+}(\delta,\delta')=\Omega^{\rm mat}_{i^+}(\delta,\delta')+\Omega^{\rm rad}_{\scri^+}(\delta,\delta').
\end{equation}
We will then use the $\Omega_{i^{+} \cup {\cal I}^{+}}$ to compute the asymptotic charge associated to super-phase rotation in section \ref{sec:superphaserotations}.
Due to the divergent (at large $r$ in null direction as well as large $t$ at fixed $r$ ) superphaserotation generator and the more relaxed fall-off behaviour of the field which is consistent with the asymptotic symmetry action, the symplectic structure has an infrared divergence and has to be regularised. We choose a specific regularisation scheme and extract a superphaserotation charge from it, 
\begin{equation}
\Omega_{i^+\cup \cal I^+}^{\rm ren}(\delta,\delta_\epsilon)=\delta Q_{+}[\epsilon].
\end{equation}

Rather remarkably, as we show, this charge is exact in the electromagnetic coupling $e$. While this property of perturbative exactness is well known and rather straightforward for the charge corresponding to leading superphaserotation, we show that it continues to hold for the charge associated to divergent superphaserotation as well. 

Our computation is done at ${\cal I}^{+} \cup i^{+}$ but an exactly analogous charge can be computed on the past boundary using antipodally identified superphaserotation generator. 

We will then show that the corresponding conservation law is independent of infrared regularisation and is in fact equivalent to the classical log soft photon theorem.

\subsection{Matter symplectic structure }
In the following we will derive the (asymptotic) symplectic structure on the space of scalar field solutions in scalar QED 
\begin{equation}
    \Omega(\delta,\delta')=\int_\Sigma dS_\mu \omega^\mu(\delta,\delta'),\quad \omega^\mu(\delta,\delta')=\delta \phi^* \nabla^\mu \delta'\phi+{\rm c.c.}-(\delta \leftrightarrow\delta').
\end{equation}
For a free complex scalar field the symplectic structure on a $\tau$ = constant hyperbolic slice $\mathcal H_\tau$ is given by
\begin{align}
\label{Omegataufree}
    \Omega^{\rm mat,free}_\tau
    &=
        \int_{\cal H_\tau} d^3y\,
        \tau^3 \omega^\tau_{\rm free}
    ,\qquad
    \omega^\tau_{\rm free}
    =
        -  \delta\phi_\text{\rm free}^* \partial_\tau\delta'\phi_\text{\rm free}
        + \text{c.c.}-(\delta \leftrightarrow \delta').
\end{align}
The free-field symplectic structure defined in terms of the vector space of free data at time-like infinity is the well-known Fock space symplectic structure, reviewed in \cite{Campiglia:2015qka}, 
\begin{equation}
\label{Omegai+free}
    \Omega^{\rm mat,free}_{i^+}= \lim_{\mathcal H_\tau \to i^+}\Omega^{\rm mat, free}_{\tau}= %
        \frac{im^2}{2(2\pi)^3}\int_{i^+} d^3y
        \left(
            \delta b^*  \delta' b
            -\delta d^* \delta' d-(\delta \leftrightarrow \delta')
        \right).
\end{equation}
In four spacetime dimensions long-range interactions dress the asymptotic free fields by `Coulombic tails'. This introduces several subtleties already in the classical theory and presents an apparent obstacle in constructing finite conserved charges. We will overcome it by implementing a suitable regularization procedure.

Our starting point is the symplectic structure of a massive complex scalar field
\begin{align}
\label{Omegatau}
    \Omega^{\rm mat}_\tau
    &=
        \int_{\cal H_\tau} d^3y\,
        \tau^3\omega^\tau
    ,\qquad
    \omega^\tau
    =
        - \delta\phi^* \partial_\tau\delta'\phi
        + \text{c.c.}-(\delta \leftrightarrow \delta')\,
\end{align}
whose late-time asymptotics accounts for long-range interactions.
By direct computation, plugging in the scalar field expansion \eqref{phiati+} with \eqref{QED_scalar_expansion}, we find the following expression for the symplectic current density for large $\tau$,
\begin{align}
\label{omegatau}
    \omega^{\tau}(\delta,\delta')
    &=
        2im
        \Big\{
            \ln\tau\Big[
                \delta \overset\ln b{}_0^* \delta' b_0
                + \delta b_0^* \delta' \overset\ln b{}_0
            \Big]+ \Big[\delta b_0^* \delta' b_0\Big]
            \nonumber\\
            &\qquad\qquad
            + \frac{\ln\tau}\tau\Big[
                \delta \overset\ln b{}_0^* \delta' b_1
                + \delta b_1^* \delta' \overset\ln b{}_0
                + \delta b_0^* \delta' \overset\ln b{}_1
                + \delta \overset\ln b{}_1^* \delta' b_0
            \Big]\\
            &\qquad\qquad
            + \frac1\tau\Big[
                \delta b_0^* \delta' b_1
                + \delta b_1^* \delta' b_0
                + \frac1{2im}\Big(
                    \delta \overset\ln b{}_0^* \delta' b_0
                    - \delta b_0^* \delta' \overset\ln b{}_0
                \Big)
            \Big]+\dots\Big\}
        \nonumber\\&\qquad\qquad
        -(\delta\leftrightarrow \delta')\nonumber
    ,
\end{align}
where we have omitted similar expressions for the negative frequency modes $\overset\ln d{}_n$ and $d_n$ which can be reinstated here and in the expressions below upon complex conjugation and replacing $m$ by $-m$. 
To distill the subtleties arising from long-range interactions let us compare \eqref{Omegatau} with its free-field counterpart \eqref{Omegataufree} and highlight several crucial points:
\begin{itemize}
    \item[(i)] {\bf Free vs interacting:} In the absence of interactions ($e\to0$) the $\tau\to \infty$ limit of \eqref{Omegatau} exists and defines the free-field symplectic structure at $i^+$ given by \eqref{Omegai+free}. The latter corresponds to the $O(\tau^0)$ term in \eqref{omegatau} which is not affected by the electromagnetic coupling. 
Indeed while the free data $b_0$ scales as $O(e^0)$ (as do the $b_n$ for $n>0$), the interacting data $\overset \ln b_0$ and $\overset \ln b_n$ for $n>0$ are of $O(e^2)$ (and similarly for $b\leftrightarrow d$) since the Coulombic modes $\overset1{A_\tau}$ and $\overset0{A_\alpha}$ in \eqref{QED_scalar_expansion}, which are sourced by the free matter current, are themselves of $O(e)$.
 \item[(ii)] {\bf Infrared divergence:} Once the electromagnetic coupling is turned on there is, compared to the free case, an `overleading' term in \eqref{Omegatau} which diverges as $\ln \tau$ in the limit $\tau \to \infty$. %
It is important to emphasize that this $\ln \tau$ divergence is simply a fact of life in four spacetime dimensions in the presence of photons: it represents the covariant phase space manifestation of the infrared divergences of scalar QED.
\item[(iii)] {\bf Tails and logs:} The coupling with the electromagnetic field leads to further corrections which decay as $\frac{\ln \tau}{\tau}$ and $\frac{1}{\tau}$ in \eqref{Omegatau}. The $\frac{1}{\tau}$ term mixes with the corresponding term in the free-field symplectic structure \eqref{Omegai+free} and the two terms are simply distinguished by their powers of the electromagnetic coupling.
At higher orders in the coupling $e$ and more subleading order in the asymptotic expansion a hierarchy of logarithmic as well as power series tails are generated which decay as $\frac{(\ln \tau)^n}{\tau^{3+k}}$ and $\frac{1}{\tau^{3+k}}$.  In appendix \ref{app:sQEDallorder} we will show that terms with $n>1$ vanish.
The power tails exhibits the mixing of free and interacting contributions, while the Coulombic interactions enter only in the log tails. 
\item[(iv)] {\bf Tree vs loop:} Ignoring long-range infrared effects as is done when matching tree-level soft theorems to charge conservation laws for leading, subleading,... superphaserotations, amounts to dropping the logarithmic tails %
which are associated to loop diagrams. We will return to this point below. 
\end{itemize}

We are now ready to derive the final expression for the IR corrected matter symplectic structure.
To deal with the $\ln \tau$ infrared divergences in the symplectic structure on $i^+$,
\begin{equation}
    \Omega_{i^+}^{\rm mat}(\delta,\delta')=\lim_{\mathcal H_\tau \to i^+} \Omega_\tau^{\rm mat}(\delta,\delta'),
\end{equation}
we introduce a late-time cutoff $\Lambda^{-1}$ and work with a suitably regularized symplectic structure. Taking one of the variation to be superphaserotations
\eqref{deltaAphi} on $i^+$ with parameter \eqref{epsi+}, we can then extract from 
\begin{equation}
\label{Omegamatreg}
\Omega^{\rm mat,reg}_{i^+}(\delta,\delta_{\bar \ep})=\delta Q^{\rm mat,reg}_{H,+}[\bar \ep].
\end{equation}
its associated Noether charge 
\begin{align}
   Q_{H,+}^{\rm mat, reg}[\bar \ep]
    &=
        -2em
        \int_{{\cal H_\tau}}d^3y\,
       \bar\ep\,\tau \,
       \Big\{
            \ln\tau\Big(
                \overset\ln b{}_0^* b_0
                + b_0^* \overset\ln b{}_0
            \Big)
            + b_0^* b_0
            \nonumber\\
            &\qquad\qquad\qquad\qquad\qquad + \frac{\ln\tau}\tau\Big(
                \overset\ln b{}_0^* b_1
                + b_1^* \overset\ln b{}_0
                + b_0^* \overset\ln b{}_1
                + \overset\ln b{}_1^* b_0
            \Big)
            \\
            &\qquad\qquad\qquad\qquad\qquad
            + \frac1\tau\Big[
                b_0^* b_1
                + b_1^* b_0
                + \frac1{2im}\left(
                    \overset\ln b{}_0^* b_0
                    - b_0^* \overset\ln b{}_0
                \right)
            \Big]
            \Big\}
            + \dots
    .\nonumber
\end{align}
Plugging in the solutions \eqref{QED_scalar_expansion} to the equations of motion, this becomes %
\begin{equation}\label{Omegamatfinal}
\badat{2}
    Q_{H,+}^{\rm mat, reg}[\bar \ep]
    &=
        2em
        \int_{{\cal H}_\tau}d^3y\,
       \Big\{
            -  \tau\,\bar\ep\, b_0^* b_0
            + \ln \tau \frac{e}{m}
            ({\cal D}^\alpha\bar\ep)
            (\p_\alpha\overset1A_\tau)b_0^*b_0
        \\&\qquad\qquad\qquad\qquad
            + \frac{1}{2im}({\cal D}^\alpha\bar\ep)
            \Big[
                b_0^* \p_\alpha b_0
                - (\p_\alpha b_0^*)  b_0+ O(e^2)
            \Big]+\dots
        \Big\}.
\eadat
\end{equation}
Interestingly, after the superphaserotation and using the equations of motion the leading divergence is linear in $\tau$. 
Its origin is the action on the matter field \eqref{deltaphi} of the superphaserotation with linearly divergent parameter \eqref{epsi+} rather than due to long-range interactions. Thus, the $O(\tau)$ divergence can be removed by means of an appropriate counterterm (see the discussion of ambiguities above equation \eqref{Oraddth} below).
At the IR cutoff $\tau=\Lambda^{-1}$ it is given by 
\begin{equation}
\label{i+counterterm}
    \Omega_{\mathcal H_\tau}^{\rm mat, c.t.}(\delta,\delta_{\bar \ep})\Big|_{\tau = \Lambda^{-1}}= -\Lambda^{-1} \int_{\mathcal H_{\tau=\Lambda^{-1}}} d^3 y\,\delta \overset3j_\tau(y)\delta_{\bar \ep} \bar \alpha(y),
\end{equation}
for some function $\bar \alpha(y)$ which on $i^+$ transforms homogeneously under superphaserotations
\begin{equation}
    \delta_{\bar \ep} \bar \alpha = \bar \epsilon,
\end{equation}
such that the counterterm symplectic structure is a total variation on field space
\begin{equation}
   \Omega_{\mathcal H_\tau}^{\rm mat, c.t.}(\delta,\delta_{\bar \ep})\Big|_{\tau = \Lambda^{-1}}=\delta Q^{\rm mat, c.t.}_{H,+}[\bar \ep].
\end{equation}
Using this and plugging in the free field current we find
\begin{align}
    Q^{\rm mat, c.t.}_{H,+}[\bar \ep]
    =
       2em \Lambda^{-1}
        \int_{\mathcal H_{\tau=\Lambda^{-1}}} d^3y\,
        \bar\ep\,
	b_0^* b_0.
\end{align}
Thus the linear late-time divergence in \eqref{Omegamatfinal} is indeed canceled upon adding the counterterm \eqref{i+counterterm}  
to the renormalized matter symplectic structure \eqref{Omegamatreg}.\footnote{Alternatively, because the coefficient of $\Lambda^{-1}$ turns out to be precisely the hard part of the conserved Noether charge associated to the leading soft photon theorem, one can argue that this term can simply be `subtracted' from the subleading asymptotic symmetry analysis as done in \cite{Campiglia:2016hvg}.
	That is, we renormalize away the linear divergence by hindsight: it is known from \cite{Campiglia:2016hvg} that this divergence is responsible for the leading Weinberg's soft photon theorem in the Ward identity.
	We could have chosen not to renormalize this term away.
    Then, in the Ward identity these terms just correspond to $\Lambda^{-1}$ times the leading soft theorem, which vanishes and we retain our results.
	This is also why we do not renormalize the logarithmic divergence: it is the logarithmic divergence that is responsible for establishing the logarithmic soft theorem.
}

This finally yields the {\it hard} matter charge on the future time-like boundary
\begin{equation}
\label{Qmati+ren}
\badat{2}
    Q^{\Lambda}_{H,+}[\bar \ep]&= Q^{\rm mat,reg}_{H,+}[\bar \ep] + Q^{\rm mat, c.t.}_{H,+}[\bar \ep]\\
    &=
\ln \Lambda^{-1}\,2e^2 \int_{i^+}d^3y\,
            ({\cal D}^\alpha\bar\ep)
            (\p_\alpha\overset1A_\tau) \big[b_0^*b_0+d_0^* d_0\big]
        \\&\quad
            - i e  \int_{i^+}d^3y\,({\cal D}^\alpha\bar\ep)
            \big[
                b_0^* \p_\alpha b_0
                - (\p_\alpha b_0^*)  b_0 +(b\leftrightarrow d^*)
                + O(e)
            \big],
\eadat
\end{equation}
where we have reinstated the negative frequency modes.
Expressing the hard charge as
\begin{equation}\label{QH+}
   Q^\Lambda_{H,+}[\bar \ep]= \ln \Lambda^{-1}Q^{(\ln)}_{H,+} [\bar \ep]+Q^{(0)}_{H,+}[\bar \ep],
\end{equation}
we can identify two distinct contributions.
The `hard log charge' is given by
\begin{equation}\label{QHln+}
\badat2
    Q^{(\ln)}_{H,+}[\bar \ep]
    &=
        \frac{e^2m}{2(2\pi)^3}\int d^3y\,
        (\calD^\alpha\bar\ep)(\calD_\alpha\overset1A_\tau)
        \left(b^*b+d^*d\right)
    ,
\eadat
\end{equation}
while the coefficient of the cutoff-independent term is given by the hard charge 
\begin{equation}\label{QH0+}
\badat2
    Q^{(0)}_{H,+}[\bar \ep]
    &=
        - \frac{iem}{4(2\pi)^3}
        \int_{i^+}d^3y\,({\cal D}^\alpha\bar\ep)
        \big[
            b^* (\p_\alpha b)
            - (\p_\alpha b^*)  b
            - d^* (\p_\alpha d)
            + (\p_\alpha d^*) d
        \big]
        + O(e^2)
    .
\eadat
\end{equation}
Here we have used $b_0=\sqrt{\frac{m}{4(2\pi)^3}}b$ to write the charges in terms of the scalar modes that satisfy the standard Poisson bracket
\begin{equation}
    i\{b(\vec p),b^*(\vec p')\}
    = i\{d(\vec p),d^*(\vec p')\}
    =(2\pi)^3(2E_p)\delta^3(\vec p-\vec p')
    .
\end{equation}
Analogous expressions are obtained at the past time-like boundary $i^-$.

\subsection{Radiative symplectic structure}
\label{sec:RadSymp}

The radiative symplectic structure of the Maxwell field has been studied extensively over decades wherein a major challenge lies in the proper treatment of boundaries and correctly identifying the radiative phase space in the presence of soft modes \cite{He:2014cra}. In the following we will derive the (asymptotic) symplectic structure of the Maxwell field in scalar QED
\begin{equation}
\label{OmegaMaxwell}
    \Omega(\delta,\delta')=\int_\Sigma dS_\mu \omega^\mu(\delta,\delta'),\quad \omega^\mu(\delta,\delta')=\delta \mathcal A_\nu \delta' \mathcal F^{\mu\nu}-(\delta \leftrightarrow\delta').
\end{equation}
On a Cauchy slice $\Sigma_t$ of constant $t$ with $dS_\mu= d^3x \,n_\mu$ and normal $n_\mu=\partial_\mu t$ the symplectic structure  is $\Omega_{\Sigma_t}=\int_{\Sigma_t} d^3x \,\omega^t $. When recast in Bondi coordinates $t=u+r$ and pushed to null infinity it becomes
$\Omega_{\scri^+}=\lim_{\Sigma_t\to \scri^+}\Omega_{\Sigma_t}=\int_{\scri^+} du d^2 \hat x \,r^2(\omega^u+\omega^r)$. 

For a photon field with `standard' large-$r$ fall-offs \eqref{calAClarger}-\eqref{calAurlarger}, and for variations that respect those fall-offs, the radiative symplectic structure %
is 
\begin{equation}
    \Omega^{\rm rad,AS}_t=\int_{\Sigma_t} d^3x \,\omega^t,\qquad \omega^t(\delta,\delta')= -\gamma^{BC}\delta A_B \partial_t\delta' A_C - (\delta \leftrightarrow \delta'),
\end{equation}
which, upon pushing the Cauchy slice $\Sigma_t$ to future null infinity,
becomes \cite{Frolov:1977bp, Ashtekar:1981bq}
\begin{equation}
\label{OmegaradASscri+}
     \Omega^{\rm rad,AS}_{\scri^+}=\lim_{\Sigma_t\to \scri^+} \Omega^{\rm rad,AS}_t=-\int_{\scri^+}  du d^2\hat x \,\gamma^{BC} \left[ \delta A_B \partial_u\delta' A_C - (\delta \leftrightarrow \delta')\right].
\end{equation}
This is the famous Ashtekar-Streubel (AS) radiative symplectic structure.\footnote{
It is by now well understood that this structure does not admit a faithful representation of the asymptotic symmetry algebra \cite{He:2014cra} because the symplectic partner of the `constant mode' ${A}^{(0)}_C$ does not exist in the radiative phase space defined in \eqref{OmegaradASscri+}. Instead the Ashtekar-Streubel symplectic structure has to be corrected by a soft symplectic pairing~\cite{He:2014cra}
\begin{align}\label{asspl}
    \Omega^\text{soft}_{{\cal I}^{+}}
   =
        \int_{S^{2}}d^2\hat x\,\gamma^{AB}
            \delta {N}^{(0)}_A\delta' C_B
        - (\delta \leftrightarrow \delta'),
\end{align}
found by He-Mitra-Porfyriadis-Strominger with the soft modes defined by
\begin{align}
    {N}^{(0)}_A(\hat x)
    &:=
        \int_{-\infty}^{\infty} du\,
        \partial_{u} A_A(u, \hat x)
    ,\qquad
    C_B
    :=
        {A}^{(0),+}_B(\hat x)
        + {A}^{(0),-}_B(\hat x)
    .
\end{align}
This correction of the radiative phase space is crucial in establishing an asymptotic symmetry interpretation for leading soft photon theorem~\cite{He:2014cra}, but will not be pertinent to our analysis. 
}
It is finite provided that the Maxwell field at early and late times falls off as
\begin{equation}
    A_C(u,\hat x)\stackrel{u\to \pm \infty}{=} A^{(0),\pm}_C(\hat x)+O\Big(\frac{1}{|u|^\#}\Big),
\end{equation} 
where $\#$ is any positive number. This includes, in particular, the asymptotic behavior \eqref{ACufalloff} involving $1/u$ tails that we found through direct analysis of the effect of the long-range interactions between photons and massive scalar matter. 
The goal of this section is to compute the radiative symplectic structure \eqref{OmegaMaxwell} for a Maxwell field with such long-range tails.
Our analysis is complicated by the appearance of Goldstone modes which do not contribute to the radiative flux but shift the `background field' by a mode linear in $r$ as well as a mode linear in $u$. Isolating their symplectic partners is rather subtle.\footnote{
Building on the prior work of Freidel and Riello \cite{Freidel:2019ohg}, this was accomplished by Peraza in \cite{Peraza:2023ivy} for the case of massless scalar QED. One may readily import this analysis to the case of massive scalar QED. In the main text we will take a more pedestrian road.}
Our starting point for the radiative symplectic structure is
\begin{equation}
\label{Omegaraddivergent}
  \Omega^{\rm rad}_{\scri^+}(\delta,\delta')=\Omega^{\rm rad,AS}_{\scri^+}(\delta,\delta')+
  \int_{\scri^+} du d^2\hat x \left[
  \left( r \delta \alpha +u \delta \beta \right) \partial_u D^C \delta'A_C
  - (\delta\leftrightarrow \delta')
  \right]
  .
 \end{equation}
To deal with the linear in $r$ divergence\footnote{Alternatively, because the $O(r)$ coefficient is precisely the soft part of the conserved Noether charge associated to the leading soft photon theorem, one can argue that this term can be `subtracted' from the subleading asymptotic symmetry analysis. The origin of this divergence, before pushing the Cauchy slice to null infinity, is a linear in $t$ term whose `subtraction' yields the same result as that in the main text \cite{Campiglia:2016hvg}.} 
we employ a by now standard covariant phase space renormalization procedure which exploits an inherent ambiguity in the definition of the symplectic structure $\omega^\mu(\delta,\delta')=\delta\Theta^\mu(\delta')-(\delta\leftrightarrow \delta')$ by a total derivative term $\Theta \mapsto \Theta+d\theta$. Subtracting the $O(r)$ divergences in \eqref{Omegaraddivergent} amounts to adding a term
\begin{equation}
    \Omega^{\rm rad,d\theta}_{\scri^+}(\delta,\delta') = \int_{\scri^+} du d^2\hat x \,(\partial_r-\partial_u)\Big[ (r\delta \alpha+u\delta \beta) D^C\delta' A_C
    - (\delta\leftrightarrow\delta')
    \Big]
    ,
    \label{Oraddth}
\end{equation}
so that the radiative symplectic structure becomes
\begin{equation}
\label{Omegaradconvergent}
\badat{2}
\Omega^{\rm rad}_{\scri^+}+\Omega^{\rm rad, d\theta}_{\scri^+} =\Omega^{\rm rad,AS}_{\scri^+}+\int_{\scri^+} du d^2\hat x \,\Big[(  \delta \alpha - \delta \beta ) D^C\delta' A_C
 - (\delta\leftrightarrow\delta')
    \Big].
\eadat
\end{equation}
Upon adding a suitable boundary term
\begin{equation}
\badat{2}
    \Omega^{\rm rad, \rhd}_{\scri^+}(\delta,\delta') &= -\int_{\scri^+} du d^2\hat x \,\partial_u\left[u (\delta \alpha-\delta \beta) D^C\delta'A_C
    - (\delta\leftrightarrow\delta')
    \right],
\eadat
\end{equation}
which is a corner term on $\scri^+$,
the linear-in-$u$ divergence can be canceled.
We finally get
\begin{equation}
\label{Omegaradreg}
\badat{2}
  \Omega^{\rm rad+d\theta+\rhd}_{\scri^+}(\delta,\delta') &=\Omega^{\rm rad}_{\scri^+}+\Omega^{\rm rad, d\theta}_{\scri^+} + \Omega^{\rm rad, \rhd}_{\scri^+}\\
    &=
    \Omega^{\rm rad,AS}_{\scri^+}
    -\left[\int_{\scri^+} du d^2\hat x \,(\delta\alpha-\delta\beta) u\partial_u D^C \delta' A_C
    - (\delta\leftrightarrow\delta')\right]
    .
\eadat
\end{equation}
Taking $\delta'$ to be the superphaserotation $\delta_\ep$ \eqref{epsscri+} whose action is \eqref{Goldstonedelta} on the Goldstone modes and $\delta_\ep A_C=0$, we obtain 
\begin{equation}
     \Omega^{\rm rad+d\theta+\rhd}_{\scri^+} (\delta,\delta_\ep)
    = -\frac{1}{2}\int_{\scri^+} du d^2\hat x \,\epsilon \,u\partial_u ( D^2 D^C \delta A_C).
\end{equation}
The integral over the null direction is still logarithmically divergent, but unlike the linear early/late retarded time divergence, this $\ln u$ divergence is due to the long-range nature of the interactions. Indeed, it is the covariant phase space manifestation on null infinity of the infrared divergences of scalar QED. As on time-like infinity we regulate the divergence at early and at late times $u\to \pm \infty$ via an infrared cutoff $\Lambda^{-1}$.
Choosing a large but finite $u_0$, we split the $u$-integral three segments $(-\Lambda^{-1},-u_0)\cup(-u_0,u_0)\cup(u_0,\Lambda^{-1})$.
Then, the Coulombic tail $\frac{1}{u}A^{(1)}_C$ in the asymptotic behavior \eqref{ACufalloff} implies that the integrals over the lower and upper segments are divergent,
\begin{equation}
\badat2
    \int_{-\Lambda^{-1}}^{+\Lambda^{-1}} du\,u\p_u A_C
    &=
        - \int_{-\Lambda^{-1}}^{-u_0} du\Big(\frac1u A^{(1)}_C\Big)
        - \int_{u_0}^{\Lambda^{-1}} du\Big(\frac1u A^{(1)}_C\Big)
        + O(\Lambda^0)
    \\ &=
        -\ln\Lambda^{-1}\Big(A^{(1),+}_C-A^{(1),-}_C\Big)
        + O(\Lambda^0)
    \\ &=
        \ln\Lambda^{-1}\int_{-\infty}^{+\infty} du\,
        \p_u(u^2\p_u A_C)
        + O(\Lambda^0)
    .
\eadat
\end{equation}
The regularized expression is a total variation,
\begin{equation}
    \Omega^{\rm rad+d\theta+\rhd,ren}_{\scri^+}(\delta,\delta_\ep)=\delta Q_{S,+}[\ep]
\end{equation}
from which we may extract the {\it soft} charge on the future null boundary
\begin{equation}
    Q_{S,+}[\ep]=\ln \Lambda^{-1} Q_{S,+}^{(\ln)}[\ep]+Q_{S,+}^{(0)}[\ep].
\end{equation}
Here the `log soft charge' is given by
\begin{equation}
\label{QSln+}
     Q_{S,+}^{(\ln)}[\ep]=-\frac{1}{2}\int_{\scri^+}du d^2\hat x \,\epsilon \,  D^2 D^C  \partial_u(u^2 \partial_u A_C)
\end{equation}
and the coefficient of the cutoff-independent term is given by the soft charge
\begin{equation}
\label{QS0+}
    Q_{S,+}^{(0)}[\ep]=-\frac{1}{2}\int_{\scri^+} du d^2\hat x \,\epsilon   \,D^2 D^C u\partial_uA_C
\end{equation}
where, in slight abuse of notation, $A_C$ now denotes the early/late time fall-offs \eqref{ACufalloff} without the Coulombic tail.
A similar analysis can be done to derive charges at the past null boundary~$\scri^-$.

We conclude this section with a few remarks.
\begin{itemize}
    \item[(i)] {\bf Free vs interacting:} %
    From the relation between the late-$u$ behavior of the Maxwell field and the late-$\tau$ behavior of the matter current \eqref{AC0+AC1intermsofj} we can read off the orders of the electromagnetic coupling $e$: while the leading free data sourced by the free matter current $A_C^{(0)}=O(e)$, the interacting matter current sources $A_C^{(1)}=O(e^3)$. 
    \item[(ii)] {\bf Infrared divergence:} The renormalized charge has a logarithmic divergence at early and late times. This is the null infinity counterpart of the logarithmic divergence found on time-like infinity and thus corresponds to the covariant phase space incarnation of the infrared divergences in scalar QED.
    \item[(i)] {\bf Tails and logs:} The $1/u$ tail of the Maxwell field originates from the leading logarithmic term in the interacting matter current. Its integral over the null boundary leads to the logarithmic divergence of the radiative symplectic structure. This is as expected and mirrors the logarithmic divergence of the matter symplectic structure which originates from the Coulombic $1/\tau$ tail. 
    \item[(i)] {\bf Tree vs loop:} Ignoring long-range infrared effects amounts to dropping the logarithmic term in the radiative symplectic structure which is associated to loop diagrams. We will get back to this point below.
\end{itemize}

\section{From charge conservation to soft photon theorems}\label{sec:sQEDcharges}
In this section we will establish the relation between the conservation laws for superphaserotation symmetries and the soft photon theorems. While for the leading soft theorem the associated superphaserotation has a finite symmetry parameter, for the logarithmic soft photon theorem we need to consider superphaserotations that diverge at large distance. Such symmetry parameters were already considered for the subleading tree-level soft photon theorem \cite{Campiglia:2019wxe} and we will, as a byproduct of our analysis, recover it here.

For the analysis of the hard charge, it will be useful to separate the physical superphaserotation modes from the charged scalar field.
We redefine the scalar field as $\phi \to e^{ie\lambda}\phi$,
such that under a superphaserotation $\bar\ep$ we have $\lambda\to \lambda+\bar\ep$ and $\phi$ stays invariant.
On the other hand, under a small gauge transformation, $\lambda$ stays invariant and $\phi$ transforms in the standard manner.
After this redefinition, the symplectic structure of the hard sector becomes
\begin{align}
    \Omega^\mu_H(\delta,\delta')
    &=
        \delta \phi^* \delta' (\nabla^\mu\phi-ieA^\mu\phi)
        + \delta \phi \delta' (\nabla^\mu\phi^*+ieA^\mu\phi^*)
        + \delta \lambda \delta' j^\mu
        - (\delta\leftrightarrow\delta')
    .
\end{align}
Since $\phi$ is invariant under a superphaserotation, the last term on the r.h.s.\ is the only one that contributes to the hard superphaserotation charges.  At $i^+$ it is readily obtained to be 
\begin{align}
    Q_{H,+}[\bar\ep]
    &=
        \int_{i^+} d^3y\, \tau^3\,\bar\ep(y) j_\tau(y)
    .
\end{align}
A similar expression can be obtained at $i^-$.

\subsection{Leading soft theorem}
Before including loops, logs and tails let us briefly review the asymptotic symmetry interpretation for the leading soft photon theorem
\begin{equation}
\lim_{\omega\to 0}\omega {\cal M}_{N+1}=
    S_{-1} {\cal M}_N.
\end{equation}
For a superphaserotation $\ep^{\scri^+}=\overset0\epsilon(\hat x)$ such that $\delta A_C(u,\hat x)=\partial_C \overset0\epsilon(\hat x)$, we get the following expression for the soft charge \cite{He:2014cra,Campiglia:2015qka} 
\begin{equation}
\label{QSleading}
     Q^{(-1)}_{S,+}[\overset0{\ep}] =\int_{S^2}   d^2\hat x \,\overset0{\ep}(\hat x) \, \mathcal J^{(-1)}_+
\end{equation}
in terms of the leading soft photon current  
\begin{equation}
   \mathcal J^{(-1)}_+\equiv \int_{-\infty}^{+\infty} du \,\partial_u D^CA_C(u,\hat x)\quad \text{with} \quad A_C\stackrel{u\to\pm\infty}=A^{(0),\pm}_C+O(\frac{1}{|u|^{\#}})
\end{equation}
and $\#$ any positive number.
Similarly, for a superphaserotation $\ep^{i^+}=\overset0{\bar{ \ep}}(y)$ such that $\delta \phi(\tau,y)=ie\overset{0}{\bar\epsilon}(y) \phi(\tau,y)$, we get the following expression for  hard charge \cite{Campiglia:2015qka}
\begin{equation}
\label{QHleading}
    Q^{(-1)}_{H,+}[\overset0{\bar{ \ep}}] =\int_{i^+} d^3 y \,\overset0{\bar{ \ep}}(y) \,\overset3j_\tau(y)
\end{equation}
with the free matter current 
\begin{equation}
    \overset{3}{j_\tau}=-\frac{em^2}{2(2\pi)^3}(b^*b-d^*d).
\end{equation}
The superscripts $(-1)$ anticipate the connection of these soft and hard charges with the leading soft photon theorem which scales as $\omega^{-1}$ in the soft expansion. Indeed, upon antipodal identification of the symmetry parameters $\overset0{\bar{ \ep}}(y)$ and $\overset0{{ \ep}}(\hat x)$ between $i^+\cup \scri^+$ and $i^-\cup \scri^-$, we obtain charges
\begin{equation}
 Q^{(-1)}_\pm=  Q^{(-1)}_{S,\pm}+ Q^{(-1)}_{H,\pm}
\end{equation}
on the future ($+$) and past ($-$) boundary that satisfy
\begin{equation}
     Q^{(-1)}_+=Q^{(-1)}_-.
\end{equation}
Upon quantization this conservation law corresponds to the leading soft photon theorem \cite{Campiglia:2015qka}.\footnote{The relation between asymptotic symmetry and soft theorem for massless scalars was found in \cite{He:2014cra}.}

\subsection{Subleading tree-level soft theorem}
\label{sec:sQEDsubtreecharges}
In \cite{Campiglia:2016hvg} it was shown that the subleading tree-level soft photon theorem
\begin{equation}
\lim_{\omega\to 0}(1+\omega \partial_\omega) {\cal M}_{N+1}=
    S_{0} {\cal M}_N
\end{equation}
can also be understood from a conservation law
\begin{equation}
     Q^{(0)}_+=Q^{(0)}_-
\end{equation}
 of charges associated to divergent superphaserotations \eqref{epsscri+} and \eqref{epsi+}. The hard and soft contributions to the future ($+$) and past ($-$) charges
\begin{equation}
 Q^{(0)}_\pm=  Q^{(0)}_{S,\pm}+ Q^{(0)}_{H,\pm}
\end{equation}
are precisely \eqref{QH0+} and \eqref{QS0+} and so we rediscovered the result of \cite{Campiglia:2016hvg}. We will now extend this result to all orders in the coupling. 

The soft charge is given by
\begin{equation}
    Q_{S,+}^{(0)}[\ep]=-\frac{1}{2}\int_{\scri^+}  d^2\hat x \,D^2\epsilon(\hat x)   \,\mathcal J^{(0)}_+
\end{equation}
in terms of the subleading soft photon current 
\begin{equation}
    \mathcal  J^{(0)}_+\equiv  \int_{-\infty}^{+\infty} du\, u\partial_u D^C A_C \quad \text{with} \quad A_C\stackrel{u\to \pm\infty}=A^{(0),\pm}_C+O(\frac{1}{|u|^{\#}})
\end{equation}
and $\#$ any positive number $>1$. It is already exact.
We can express, using the all-orders derivation of Appendix \ref{app:sQEDallorder}, the hard charge as
\begin{equation}
\label{Q0H+allorder}
    Q^{(0)}_{H,+}[\bar \ep]=%
    \int_{i^+}d^3y\,\bar\ep(y)\,  \overset4j_\tau(y),
\end{equation}
where the current receives contributions from both free and interaction terms
\begin{align}
    \overset4j_\tau
    &=
    \frac{iem}{4(2\pi)^3}(
        b^*\calD^2b
        -b\calD^2b^*
        -d^*\calD^2d
        + d\calD^2d^*
    )
    \nonumber\\&\quad
    + \frac{e^2m}{2(2\pi)^3}
    \Big(
        2\overset1A_\tau
        - \calD^2\overset1A_\tau
        + (\overset0A_\alpha-\calD_\alpha\overset1A_\tau) \calD^\alpha
    \Big)(b^*b+d^*d)
    .
\end{align}
The hard charge \eqref{Q0H+allorder} receives no further corrections beyond $O(e^3)$ and is thus exact.
Recall that due to the appearance of logarithms these subleading charges are, however, ambiguous. The logarithmic soft theorem, which is universal and unambiguous, is what we seek to give an asymptotic symmetry interpretation.

\subsection{Logarithmic soft theorem}
\label{sec:sQEDlogcharges}

Our goal is to identify a conservation law of charges derived from a first principles covariant phase space approach which is equivalent to the classical logarithmic soft photon theorem 
\begin{equation}
  \lim_{\omega \to 0}\partial_\omega \omega^2 \partial_\omega {\cal M}_{N+1}=
    S^{(\ln \omega)}_{0,{\rm classical}} {\cal M}_N.
\end{equation}
We expect the latter to be associated with the same divergent superphaserotations \eqref{epsscri+} and \eqref{epsi+} as the tree-level subleading soft photon theorem but with an infrared-corrected Noether charge given by our logarithmic soft and hard charges \eqref{QSln+} and \eqref{QHln+}. The soft log charge is given by 
\begin{equation}
     Q_{S,+}^{(\ln)}[\ep]=-\frac{1}{2}\int_{S^2} d^2\hat x \,D^2\epsilon(\hat x) \,    \mathcal J^{(\ln)}_+
\end{equation}
in terms of the `logarithmic soft photon current' 
\begin{equation}
    \mathcal J^{(\ln)}_+\equiv \int_{-\infty}^{+\infty} du\,\partial_u(u^2 \partial_u D^CA_C)\quad \text{with} \quad A_C\stackrel{u\to \pm\infty}=A^{(0),\pm}_C+\frac{1}{u}A^{(1),\pm}_C+O(\frac{1}{|u|^{1+\#}})
\end{equation}
and $\#$ any positive number, and is exact.
The hard log charge can be expressed as
\begin{equation}
\label{QHln+intj}
  Q^{(\ln)}_{H,+}[\bar \ep]%
  =\int_{i^+} d^3y \,\bar\ep(y) \,\overset{4,\ln}{j_\tau}(y),
\end{equation}
with the interacting current given
\begin{equation}
    \overset{4,\ln}{j_\tau}
    =
        -\frac{e^2m}{2(2\pi)^3}
            \calD_\alpha\Big((\calD^\alpha\overset1A_\tau)(b^*b+d^*d)\Big)
\end{equation}
This expression is exact to all orders in $e$ as we show in Appendix \ref{app:sQEDallorder}.

Upon antipodal identification of the symmetry parameters $\bar{ \ep}(y)$ and ${ \ep}(\hat x)$ between $i^+\cup \scri^+$ and $i^-\cup \scri^-$, we obtain charges
\begin{equation}
 Q^{(\ln)}_\pm=  Q^{(\ln)}_{S,\pm}+ Q^{(\ln)}_{H,\pm}
\end{equation}
on the future ($+$) and past ($-$) boundary satisfying
\begin{equation}
    Q^{(\ln)}_+=Q^{(\ln)}_-
\end{equation}
This classical conservation law was engineered in \cite{Campiglia:2019wxe} from the classical logarithmic soft photon theorem. Here we derived it from first principles and showed that it holds to all orders in the coupling $e$. Our covariant phase space result thus exhibits the same type of universality as the (classical) logarithmic soft photon theorem and unequivocally establishes its symmetry interpretation.

\section{Conclusion}
Just as the S-matrix is infrared divergent in scalar QED, in the classical theory, the symplectic structure at the boundary also suffers from late-time divergences and has to be regularised. 
In the context of scalar QED, a regularised S-matrix can be used to define infrared-safe observables such as inclusive or semi-inclusive cross sections.
In the classical theory we have shown that the regularised Noether charge defines an infrared-safe asymptotic conservation law which is equivalent to the classical logarithmic soft theorem.
In the S-matrix computation this amounts to going beyond the tree-level approximation and include the relevant loop diagrams that survive in the classical limit. This is mirrored in the covariant phase space analysis by accounting for late-time tails and considering divergent superphaserotations. Here we focused on the classical logarithmic soft photon theorem in scalar QED which is one-loop exact and requires us to account for the leading effects of the $1/\tau$ and $1/u$ tails as well as consider $O(\tau)$ and $O(r)$ divergent superphaserotations. While we have given a first principle derivation of the classical conservation law which is equivalent to the (classical) log soft theorem, in \cite{Campiglia:2019wxe, AtulBhatkar:2019vcb} it was argued that the quantisation of this conservation law implies the quantum log soft photon theorem in scalar QED. It would be interesting to revisit this quantisation in light of the current work.

\acknowledgments

We would like to thank Prahar Mitra and Ashoke Sen for discussions. AL would also like to thank Miguel Campiglia for discussions and collaboration on related projects. 
SC and AP are supported by the European Research Council (ERC) under the European Union’s Horizon 2020 research and innovation programme (grant agreement No 852386). 
This work was supported by the Simons Collaboration on Celestial Holography. This research was supported in part by Perimeter Institute for Theoretical Physics. Research at Perimeter Institute is supported by the Government of Canada through the Department of Innovation, Science and Economic Development and by the Province of Ontario through the Ministry of Research, Innovation and Science.

\newpage

\appendix

\section{Tails from time-like to null infinity}\label{app:ACufalloff}

The $u\to \infty$ behavior of the Cartesian components of the photon field \eqref{Amuru} at $\scri^+$ are 
\begin{equation}
\badat{3}\label{Alogu0}
    \overset0{A_\mu}(x^A)&=-\frac{1}{4\pi}\int d^3y (-q\cdot \Y)^{-1} \Y_\mu \overset3{j_\tau},\\
     \overset{1,\ln}{A_\mu}(x^A)&=\frac{1}{4\pi}\int d^3y\left(-\Y_\mu \overset{4,\ln}{j_\tau}+\mathcal D^\alpha \Y_\mu \overset{3,\ln}{j_\alpha}\right)\\
    \overset1{A_\mu}(x^A)&=\frac{1}{4\pi}\int d^3y \left[\left(-\Y_\mu \overset4{j_\tau}+\mathcal D^\alpha \Y_\mu \overset3{j_\alpha}\right)-\ln(-q\cdot \Y) \left(-\Y_\mu\overset{4,\ln}{j_\tau} +\mathcal D^\alpha \Y_\mu \overset{3,\ln}{j_\alpha}\right)\right].
\eadat
\end{equation} 
Now we use current conservation $\nabla^\mu j_\mu=0$, which in hyperbolic components translates to 
\begin{equation}
\label{currentconservation}
    -\left(\partial_\tau+\frac{3}{\tau}\right) j_\tau +\frac{1}{\tau^2} \mathcal D^\alpha j_\alpha=0.
\end{equation}
For the relevant terms in the $1/\tau$ expansion we have $\overset{4,\ln}{j_\tau}+\mathcal D^\alpha\overset{3,\ln}{j_\alpha}=0$ and $\overset{4}{j_\tau}-\overset{4,\ln}{j_\tau}+\mathcal D^\alpha\overset{3}{j_\alpha}=0$.\footnote{Note that while \eqref{currentconservation} together with the fall-offs \eqref{jtaualphai+} imply $\overset{4,\ln}{j_\tau}+\mathcal D^\alpha\overset{3,\ln}{j_\alpha}=0$ at $O(e^3)$, we will show in Appendix \ref{app:sQEDallorder} that this actually holds to all orders.}
As a consequence the $\ln u/u$ term takes the form of an integral over a total derivative which vanishes \cite{Campiglia:2019wxe}
\begin{equation}
     \overset{1,\ln}{A_\mu}(x^A)=\frac{1}{4\pi}\int d^3y\mathcal D^\alpha \left(\Y_\mu \overset{3,\ln}{j_\alpha}\right)=0.
\end{equation}
The $1/u$ term simplifies to
\begin{equation}
\badat{2}
      \overset1{A_\mu}(x^A)&=\frac{1}{4\pi}\int d^3y \left[\Y_\mu \mathcal D^\alpha \overset{3,\ln}{j_\alpha}+\mathcal D^\alpha\left(\Y_\mu\overset3{j_\alpha}\right)- \ln(-q\cdot \Y) \mathcal D^\alpha\left(\Y_\mu \overset{3,\ln}{j_\alpha}\right)\right]\\
    &=\frac{1}{4\pi}\int d^3y (q\cdot \Y)^{-1}
    q^\nu J^\alpha_{\mu\nu}(y)\overset{3,\ln}{j_\alpha}(y),
\eadat
\end{equation}
where $J^\alpha_{\mu\nu}\equiv \Y_\mu \mathcal D^\alpha \Y_\nu-\Y_\nu \mathcal D^\alpha \Y_\mu$ and we used integration by parts.

\section{Logarithmic charge to all orders in the coupling}
\label{app:sQEDallorder}
In this appendix, we show that the expressions \eqref{QHln+} and  \eqref{QSln+} for the logarithmic hard and soft charges in scalar QED are one-loop exact, in the sense that they do not receive further corrections at higher orders in the coupling constant $e$.
For this purpose, we solve the scalar equation of motion without dropping any terms. %

To incorporate terms of higher power in the coupling constant, we allow for terms with higher powers of $\ln\tau$ in the large-$\tau$ expansion of $\phi$.
\begin{align}\label{Ataualphai+higherlogs}
    \phi(\tau,y)
    &=
        \sum_{k=0}^\infty \sum_{n=0}^\infty \frac{(\ln\tau)^n}{\tau^{\frac32+k}} \Big(e^{-im\tau} b_{k,n}(y) + e^{im\tau} d^*_{k,n}(y)\Big)
    .
\end{align}
Matching with the notation used in the main text, we shall also refer to the first few coefficients as $b_k\equiv b_{k,0}$ and $\overset\ln b_k\equiv b_{k,1}$.
Since we are solving linear equations, we solve for the positive-frequency modes $b_{k,n}$ separately, and then obtain the results for negative-frequency modes $d_{k,n}^*$ by taking $m\to-m$.
For the gauge field we take the falloff
\begin{align}\label{Ataualphai+}
\badat2
    \calA_\tau(\tau,y)
    &=
        \frac1\tau \overset1A_\tau(y)
        + \frac{\ln\tau}{\tau^2}\overset\ln A_\tau(y)
        + \frac{1}{\tau^2}\overset2A_\tau(y)
        + \cdots
    ,\\
    \calA_\alpha(\tau,y)
    &=
        \overset0A_\alpha(y)
        + \cdots
    .
\eadat
\end{align}
The gauge field does not admit corrections of the form $\tau^{-2}(\ln\tau)^n$ with $n\geq 2$ at higher order in the coupling.
We will see later in this section that the terms in the current $j_\tau$ which fall off as $\tau^{-4}(\ln\tau)^n$ with $n\geq 2$ vanish; these are the terms that source the higher powers of $\ln\tau$ terms in the gauge field via Maxwell equations.

Plugging \eqref{Ataualphai+higherlogs} and \eqref{Ataualphai+} into the scalar equation of motion \eqref{sQEDphieom} yields
\begin{align}
	0
	&=
		\sum_{n=0}^\infty \frac{(\ln\tau)^n}{\tau^{5/2}} {\rm Eq}_{\frac52,n}
		+ \sum_{n=0}^\infty \frac{(\ln\tau)^n}{\tau^{7/2}} {\rm Eq}_{\frac72,n}
		+ \cdots
    .
\end{align}
Each ${\rm Eq}_{\alpha,n}$ should independently vanish.
At leading order,
\begin{align}
	0
	&\stackrel{!}{=}
		{\rm Eq}_{\frac52,n}
	=
		2me^{-im\tau}\left(
			e\overset1A_\tau b_{0,n} + i(n+1) b_{0,n+1}
		\right)
	,
\end{align}
which determines all $b_{0,n+1}$ in terms of $b_{0,0}$ as
\begin{align}
	b_{0,n} = \frac{1}{n!}(ie\overset1A_\tau)^n b_{0,0}
	.
	\label{b0n}
\end{align}
This implies that the series of $(\ln\tau)^n\tau^{-3/2}$ terms in the large-$\tau$ expansion of the scalar field exponentiates to a phase,
\begin{align}
	\phi
	&=
		\frac{e^{ie\ln\tau \overset1A_\tau}}{\tau^{3/2}}e^{-im\tau} b_{0,0}
		+ \sum_{n=0}^\infty \frac{(\ln\tau)^n}{\tau^{5/2}}e^{-im\tau}b_{1,n}
		+ \dots\,
	.
\end{align}
The equation that determines $b_{1,n}$ is 
\begin{align}
    0\stackrel{!}{=}{\rm Eq}_{\frac72,n}
    &=
        e^{-im\tau}\Bigg[
        \left(
            {\cal D}^2
            + \frac34
            + e^2(\overset1A_\tau)^2
            - e^2(\overset0A_\alpha)^2
            - 3ie\overset1A_\tau
            + 2em \overset2A_\tau
            - 2ie \overset0A_\alpha{\cal D}^\alpha
        \right)b_{0,n}
        \nonumber\\&\quad
        + 2em \overset\ln A_\tau b_{0,n-1}
        + (n+1)(1+2ie\overset1A_\tau)b_{0,n+1}
        - (n+1)(n+2)b_{0,n+2}
        \nonumber\\&\quad
        - 2im(1+ie\overset1A_\tau) b_{1,n}
        + 2im(n+1)b_{1,n+1}
        \Bigg]
    ,
\end{align}
where we employ the shorthand notation $(\overset0A_\alpha)^2\equiv k^{\alpha\beta} \overset0A_\alpha \overset0A_\beta$. 
For the special case $n=0$ we define $b_{0,n-1}=b_{0,-1}\equiv 0$.
Plugging in \eqref{b0n}, this simplifies to
\begin{align}
	0
	&=
		\left(
			{\cal D}^2
			+ \frac34
			- e^2(\overset0A_\alpha)^2
			- 2ie\overset1A_\tau
			+ 2em \overset2A_\tau
			- 2ie \overset0A_\alpha{\cal D}^\alpha
		\right)\left(\frac{1}{n!}(ie\overset1A_\tau)^nb_{0,0}\right)
		\nonumber\\&\quad
		+ 2em \overset\ln A_\tau b_{0,n-1}
		- 2im(1+ie\overset1A_\tau) b_{1,n}
		+ 2im(n+1)b_{1,n+1}
    .
\end{align}
Rearranging terms to solve for $b_{1,n}$ we find
\begin{align}
	b_{1,n}
 &=\frac{1}{1+ie\overset1A_\tau}\Big[
		 (n+1)b_{1,n+1}- ie \overset\ln A_\tau b_{0,n-1}
		\nonumber\\&\quad
 -\frac{i}{2m}
		\left(
			{\cal D}^2
			+ \frac34
			- e^2(\overset0A_\alpha)^2
			- 2ie\overset1A_\tau
			+ 2em \overset2A_\tau
			- 2ie \overset0A_\alpha{\cal D}^\alpha
		\right)\bigg(\frac{(ie\overset1A_\tau)^n}{n!} b_{0,0}\bigg)
		\Big].
\end{align}
Here the factor $(1+ie\overset1A_\tau)^{-1}$ is a formal expression that represents the infinite series
\begin{align}
	\frac{1}{1+ie\overset1A_\tau}
	&=
		\sum_{l=0}^\infty (-ie\overset1A_\tau)^l
\end{align}
The expression for $b_{1,n}$ is given in terms of $b_{1,n+1}$, since we already know $b_{0,n-1}$ from \eqref{b0n}.
This is a recurrence relation whose solution can be written as an infinite series:
\begin{align}
    b_{1,n}
    &=
        \frac{-i}{2mn!}\sum_{r=n}^\infty 
        \frac{1}{(1+ie\overset1A_\tau)^{r-n+1}}
        \left(
            \calD^2
            + \frac34
            - e^2(\overset0A_\alpha)^2
            - 2ie\overset1A_\tau
            + 2em \overset2A_\tau
            - 2ie \overset0A_\alpha\calD^\alpha
        \right)\left((ie\overset1A_\tau)^rb_0\right)
        \nonumber\\&\quad
        - \frac{ie}{n!}\overset\ln A_\tau b_0\sum_{r=n}^\infty 
        \frac{r (ie\overset1A_\tau)^{r-1}}{(1+ie\overset1A_\tau)^{r-n+1}}
    .
\end{align}
After acting with ${\cal D}_\alpha$ and ${\cal D}^2$ on $(ie\overset1A_\tau)^r b_0$, we notice that every series re-sums to powers of $(ie\overset1A_\tau)$ or $(1+ie\overset1A_\tau)$, leading to the following simple closed form:
\begin{align}\label{b1nexact}
    b_{1,n}
    &=
        \frac{ie^2}{2mn!}
            n(n-1)(ie\overset1A_\tau)^{n-2}
            (\calD_\alpha\overset1A_\tau)(\calD^\alpha\overset1A_\tau)b_0
        \nonumber\\&\quad
        + \frac{e}{mn!}
            \left[
                (ie\overset1A_\tau)^n
                + n(ie\overset1A_\tau)^{n-1}
            \right]
            \left[
                (\calD^\alpha \overset1A_\tau)
                    \left(
                        \calD_\alpha
                        + ie (\calD_\alpha\overset1A_\tau)
                        - ie \overset0A_\alpha
                    \right)
                + \frac12 (\calD^2\overset1A_\tau)
                - i m \overset\ln A_\tau
            \right]b_0
        \nonumber\\&\quad
        - \frac{i}{2mn!}(ie\overset1A_\tau)^n
            \left(
                {\cal D}^2
                +\frac34
                - 2ie\overset0A_\alpha \calD^\alpha
                - e^2(\overset0A_\alpha)^2
                - 2ie\overset1A_\tau
                + 2em \overset2A_\tau
            \right)b_0
\end{align}
{\it This expression is exact to all orders in $e$.} \\

The r.h.s.\ of \eqref{b1nexact} depends on  $\overset1A_\tau$, $\overset0A_\alpha$, $\overset\ln A_\tau$ and $\overset2A_\tau$. some of these components may depend on the modes $b_{1,m}$ for $m\, < n$. In which case, eqn. \eqref{b1nexact} is an identity involving $b_{1,n}$. However, this turns out not to be true. And the gauge modes that appear on the r.h.s.\ of this equation can be expressed in terms of just free data $b_0$.
To see this, we consider the Maxwell's equations relevant for these gauge modes
\begin{align}
    \overset3j_\tau
    = -\calD^2 \overset1A_\tau
    ,\qquad
    (\calD^2+2)\overset0A_\alpha
    = 2\calD_\alpha\overset1A_\tau
    ,\\
    \overset{4,\ln}{j_\tau}
    = -(\calD^2+1)\overset\ln A_\tau
    ,\qquad
    \overset4j_\tau
    = -(\calD^2+1)\overset2A_\tau
    .
\end{align}
The two equations in the first line imply that $\overset1A_\tau$ is sourced by the free current $\overset3j_\tau$ which contains only $b_0$, and $\overset0A_\alpha$ in turn is sourced by $\overset1A_\tau$. Thus, both $\overset1A_\tau$ and $\overset0A_\alpha$ are given in terms of $b_0$.
The second line tells us that $\overset\ln A_\tau$ and $\overset2A_\tau$ are sourced by $\overset{4,\ln}{j_\tau}$ and $\overset4j_\tau$ respectively.
We show in \eqref{j4nt_allorder} that these two modes of the current have closed-form expressions given just in terms of $b_0$, $\overset1A_\tau$ and $\overset0A_\alpha$.
It follows that all terms on the r.h.s.\ of \eqref{b1nexact} can be expressed just in terms of $b_0$ and \eqref{b1nexact} hence is a solution for $b_{1,n}$.

In the rest of this appendix, we will make use of this result to show the all-order exactness of the soft and hard charges corresponding to the superphaserotations \eqref{epsscri+} and \eqref{epsi+}.

\subsection*{Logarithmic hard charge}
The log hard charge is given by
\begin{align}\label{QHln+intj_app}
    Q^{(\ln)}_{H,+}[\bar \ep]
    =
        \int_{i^+} d^3y\,
        \bar\ep(y) \overset{4,\ln}{j_\tau}(y)
    ,
\end{align}
where $\overset{4,\ln}{j_\tau}$ is the coefficient of $\tau^{-4}\ln\tau$ of the current as given in \eqref{jtaualphai+}.
Plugging in the expansions \eqref{Ataualphai+higherlogs} and \eqref{Ataualphai+}, we obtain the following expression,
\begin{equation}
	\overset{4,\ln}{j_\tau}
	=
        -2ie(b_{0,2} - i e \overset1A_\tau \overset\ln b_0) b_0^*
        - 2me (\overset\ln b_0b_1^* + b_0\overset\ln b{}_1^*)
		+ \text{c.c.}~
    .
\end{equation}
Equation \eqref{b0n} implies that the first term on the r.h.s.\ is purely imaginary,
\begin{align}
	-2ie(b_{0,2}-ie\overset1A_\tau b_{0,1})b_{0,0}^*
	&=
		-ie^3 (\overset1A_\tau)^2 |b_0|^2
	,
\end{align}
and thus is canceled out by its complex conjugate.
Using \eqref{b0n} and \eqref{b1nexact}, we obtain the following expression,
\begin{align}
	\overset\ln b_1b_0^*
	+ b_1\overset\ln b{}_0^*
	&=
		\frac{e}{2m}({\cal D}^2\overset1A_\tau)|b_0|^2
		+ \frac{e}{m}({\cal D}^\alpha\overset1A_\tau)
			({\cal D}_\alpha b_0)b_0^*
		\nonumber\\&\quad
		+ ie\left[
			em({\cal D}^\alpha\overset1A_\tau)({\cal D}_\alpha\overset1A_\tau)
			- em\overset0A_\alpha ({\cal D}^\alpha\overset1A_\tau)
			- \overset\ln A_\tau
		\right]|b_0|^2
	.
\end{align}
Note that the second line is purely imaginary and hence get canceled out by adding its complex conjugate.
The remaining terms can be organized into a total derivative:
\begin{align}
    \overset\ln b_1b_0^*
    + b_1\overset\ln b{}_0^*
    + \text{c.c.}
    &=
        \frac em{\cal D}^\alpha\left[
            ({\cal D}_\alpha\overset1A_\tau)|b_0|^2
        \right]
    .
\end{align}
Therefore, after putting in the contribution from negative-frequency modes, we obtain the current $\overset{4,\ln}{j_\tau}$ to be
\begin{align}
    \overset{4,\ln}{j_\tau}
    &=
		-2e^2{\cal D}^\alpha\left[
			({\cal D}_\alpha\overset1A_\tau)(|b_0|^2+|d_0|^2)
		\right]
	.
\end{align}
Using $b_0(y) = \frac{\sqrt m}{2(2\pi)^{3/2}}b(y)$ and integrating by parts, we obtain the following expression for the log hard charge 
\begin{align}
    Q^{(\ln)}_{H,+}[\bar \ep]
    =
        \frac{e^2m}{2(2\pi)^3}
        \int_{i^+} d^3y\,
        ({\cal D}^\alpha\bar\ep)
			({\cal D}_\alpha\overset1A_\tau)(b^*b + d^*d)
    .
\end{align}
This is in exact agreement with the expression \eqref{QHln+} in the main text.
This proves that the log hard charge is exact at $O(e^3)$ and receives no corrections at higher orders.

\subsection*{Subleading `tree-level' hard charge}

The subleading tree-level hard charge can be expressed in terms of the coefficient $\overset4j_\tau$ of the current \eqref{jtaualphai+} as
\begin{align}\label{Q0Hi+_app}
    Q^{(0)}_{H,+}[\bar\ep]
    &=
        \int_{i^+}d^3y\,
       \, \bar\ep(y)\,
        \overset4j_\tau(y)
        .
\end{align}
Using the falloff \eqref{Ataualphai+higherlogs}, we find this coefficient to be
\begin{align}
    \overset4j_\tau
    &=
        -e^2\overset1A_\tau (|b_0|^2+|d_0|^2)
        + ie ( b_0\overset\ln b{}_0^* + d_0^* \overset\ln d{}_0)
        - 2em (b_1b_0^*-d_0d_1^*)
        + \text{c.c.}~
    .
\end{align}
Plugging in \eqref{b0n} and \eqref{b1nexact}, this becomes
\begin{align}
    \overset4j_\tau
    &=
    ie(
        b_0^*\calD^2b_0-b_0\calD^2b_0^*
        + d_0\calD^2d_0^*-d_0^*\calD^2d_0
    )
    \nonumber\\&\quad
    + 2e^2
    \Big(
        2\overset1A_\tau
        - \calD^2\overset1A_\tau
        + (\overset0A_\alpha-\calD_\alpha\overset1A_\tau) \calD^\alpha
    \Big)(|b_0|^2+|d_0|^2)
    .
\end{align}
Thus, the tree-level hard charge \eqref{Q0Hi+_app} is
\begin{align}
    Q^{(0)}_{H,+}[\bar\ep]
    &=
        \frac{iem}{4(2\pi)^3}
        \int_{i^+}
        d^3y\,
        \bar\ep(y)\left(
            b^*\calD^2b
            - b\calD^2b^*
            + d\calD^2d^*
            - d^*\calD^2d
        \right)
        \nonumber\\&\quad
        + \frac{e^2m}{2(2\pi)^3}
        \int_{i^+} d^3y\,
        \bar \ep(y)
        \Big(
            2\overset1A_\tau
            - \calD^2\overset1A_\tau
            + (\overset0A_\alpha-\calD_\alpha\overset1A_\tau) \calD^\alpha
        \Big)(b^*b+d^*d),
\end{align}
where we have used $b_0=\frac{\sqrt m}{2(2\pi)^{3/2}}b$. This corrects the expression \eqref{QH0+} which was computed perturbatively in $e$ to leading order.

\subsection*{Logarithmic soft charge }

Higher powers of $\ln\tau$ in the current imply that we should consider a generalization of the large-$u$ expansion of the gauge field \eqref{Amuru},
\begin{align}\label{Amuru_generalization}
    \calA_\mu(x)
    &=
        \frac1r\left[
            \overset0A_\mu(x^A)
            + \sum_{n=1}^\infty \frac{(\ln u)^n}{u}\overset{1,n}{A_\mu}(x^A)
            + \frac1u \overset1A_\mu(x^A)
            + \cdots
        \right]
\end{align}
where
\begin{align}
    \overset{1,n}{A_\mu}(x^A)
    &=
        \frac{1}{4\pi}
        \int d^3y
        \sum_{r=n}^\infty
        \binom{r}{n}\big(-\ln(-q\cdot \Y)\big)^r
        \left(
            -\Y_\mu \overset{4,r}{j_\tau}
            + \mathcal D^\alpha \Y_\mu \overset{3,r}{j_\alpha}
        \right)
        ,\qquad n\geq 1
        .
\end{align}
In what follows, we show that $\overset{4,r}{j_\tau}$ and $\overset{3,r}{j_\alpha}$ are both zero for all $r\geq 2$ (see \eqref{jtvanish} and \eqref{javanish}), which immediately implies that
\begin{equation}
\badat3
    \overset{1,0}{A_\mu}(x^A)
    &=
        \frac{1}{4\pi}
        \int d^3y
        \left[
        \left(
            -\Y_\mu \overset{4,0}{j_\tau}
            + \mathcal D^\alpha \Y_\mu \overset{3,0}{j_\alpha}
        \right)
        - \ln(-q\cdot\Y)
        \left(
            -\Y_\mu \overset{4,1}{j_\tau}
            + \mathcal D^\alpha \Y_\mu \overset{3,1}{j_\alpha}
        \right)
        \right]
    ,\\
    \overset{1,1}{A_\mu}(x^A)
    &=
        \frac{1}{4\pi}
        \int d^3y
        \left(
            -\Y_\mu \overset{4,1}{j_\tau}
            + \mathcal D^\alpha \Y_\mu \overset{3,1}{j_\alpha}
        \right)
    =
        -\frac{1}{2\pi}
        \int d^3y\,\Y_\mu\, \overset{4,2}{j_\tau}
    =
        0
    ,\\
    \overset{1,n}{A_\mu}(x^A)
    &=
        0
        \qquad\text{for all $n\geq 2$}
    ,
\eadat
\end{equation}
which agrees with \eqref{Alogu0}.
In the second line, we have used partial integration followed by the current conservation \eqref{currentconservation},
\begin{equation}
    \overset{4,n}{j_\tau}+  \mathcal D^\alpha \overset{3,n}{j_\alpha} =(n+1)\overset{4,n+1}{j_\tau}
    \ ,\qquad n\geq 0
    .
\end{equation}
to write the integrand as $\overset{4,2}{j_\tau}$ which vanishes.
$\overset{1,n}{A_\mu}=0$ for $n\geq1$ to all orders in the coupling: the expansion \eqref{Amuru_generalization} does not admit corrections of the form $\frac{(\ln u)^n}{u}$ with $n\geq 1$ even at higher powers in the coupling.

The crucial ingredient to this argument is that
\begin{equation}
    \overset{4,n}{j_\tau} = 0
    ,\qquad
    \overset{3,n}{j_\alpha} = 0
    ,\qquad
    n\geq 2,
\end{equation}
to all orders in the coupling, which we now demonstrate.

We first obtain a closed-form expression for the current $\overset{4,n}{j_\tau}$ for $n\geq 0$.
By direct computation using the expansion \eqref{Ataualphai+higherlogs}, we obtain the following expression for the current at the order $\tau^{-4}(\ln\tau)^n$,
\begin{align}
	\overset{4,n}{j_\tau}
	&=
		-2e^2 \overset1A_\tau \sum_{k=0}^n b_{0,k}b_{0,n-k}^*
		\nonumber\\&\quad+ ie \sum_{k=0}^{n+1}
			(n+1-2k)b_{0,k}b_{0,n+1-k}^*
		\\ &\quad-2em \sum_{k=0}^n
		\left(
			b_{1,k}b_{0,n-k}^*
			+ \text{c.c.}
		\right)
    ~.
    \nonumber
	\label{jln4}
\end{align}
There are three series that appear.
It turns out that the first two cancel each other out, and that the last one only contains terms with either $\delta_{n,1}$ or $\delta_{n,0}$.
We examine them one by one:
\begin{itemize}
\item {\bf First series:}
Plugging in the expression \eqref{b0n}, the first series becomes 
\begin{align}
    \sum_{k=0}^n b_{0,k}b_{0,n-k}^*
    &=
        \frac{|b_0|^2}{n!}(ie\overset1A_\tau)^n\sum_{k=0}^n \binom{n}{k} (-1)^{n-k}
    .
\end{align}
This is a binomial expansion for $(1-1)^n=\delta_{n,0}$.
Thus,
\begin{align}
    \sum_{k=0}^n b_{0,k}b_{0,n-k}^*
    &=
        |b_0|^2 \delta_{n,0}
    .
\end{align}
\item {\bf Second series:}
Writing out the second term using \eqref{b0n} yields
\begin{align}
    \sum_{k=0}^{n+1}
        (n+1-2k)b_{0,k}b_{0,n+1-k}^*
    &=
        (-ie\overset1A_\tau)^{n+1}|b_0|^2
        \sum_{k=0}^{n+1}
        (-1)^{k}\frac{(n+1-2k)}{k!(n+1-k)!}
    \label{gb01}
    .
\end{align}
The sum on the r.h.s.\ can be rewritten as
\begin{align}
    \sum_{k=0}^{n+1}(-1)^k\frac{(n+1-2k)}{k!(n+1-k)!}
    &=
        \frac{1}{n!}\sum_{k=0}^{n+1}\binom{n+1}{k}(-1)^k
        + \frac{2}{n!}\sum_{k=0}^{n}\binom{n}{k}(-1)^k
    .
\end{align}
The first sum on the r.h.s.\ is the binomial expansion of $(1-1)^{n+1}=\delta_{n+1,0}$; this is identically zero since $n$ is non-negative.
The second sum is $\delta_{n,0}$, so we conclude that
\begin{align}
    \sum_{k=0}^{n+1}(n+1-2k)b_{0,k}b^*_{0,n+1-k}
    &=
        -2ie\overset1A_\tau |b_0|^2\delta_{n,0}
        .
\end{align}

\item {\bf Third series:}
We are left with the last series on the r.h.s.\ of \eqref{jln4},
\begin{align}
        \sum_{k=0}^n
        \left(
            b_{1,k}b_{0,n-k}^*
            + \text{c.c.}
        \right)
    .
\end{align}
Plugging in the expressions \eqref{b0n} and \eqref{b1nexact}, we find that every sum becomes a binomial expansion that reduces to either $\delta_{n,0}$ or $\delta_{n,1}$:
\begin{align}
    \sum_{k=0}^n
    \left(
        b_{1,k}b_{0,n-k}^*
        + \text{c.c.}
    \right)
    &=
        - \frac{\delta_{n,0}}{2m}
            \left[
                i \left(
                    b_0^*{\cal D}^2b_0
                    - b_0{\cal D}^2b_0^*
                \right)
                + 2e\overset0A_\alpha \left(
                    b_0^*\calD^\alpha b_0
                    + b_0\calD^\alpha b_0^*
                \right)
                + 4e\overset1A_\tau |b_0|^2
            \right]
        \nonumber\\&\quad
        + \frac{e}{m}
            (\delta_{n,0}+\delta_{n,1})
            \calD_\alpha(\calD^\alpha\overset1A_\tau)|b_0|^2
\end{align}

\end{itemize}
Collecting the results and plugging them into \eqref{jln4}, we obtain
\begin{align}\label{j4nt_allorder}
    \overset{4,n}{j_\tau}
    &=
        -2e^2
            (\delta_{n,0}+\delta_{n,1})
            \calD_\alpha\Big((\calD^\alpha\overset1A_\tau)|b_0|^2\Big)
        \nonumber\\&\quad
        + \delta_{n,0}
            \left[
                ie \left(
                    b_0^*{\cal D}^2b_0
                    - b_0{\cal D}^2b_0^*
                \right)
                + 2e^2\overset0A_\alpha \calD^\alpha|b_0|^2
                + 4e^2\overset1A_\tau |b_0|^2
            \right]
        + (b\to d^*)
    .
\end{align}
Therefore, we have established that
\begin{equation}\label{jtvanish}
    \overset{4,n}{j_\tau}=0
    \qquad\text{ for all $n\geq 2$}
    .
\end{equation}
Next, we show that $\overset{4,n}{j_\alpha}=0$ for all $n\geq 2$.
Using the expansion \eqref{Ataualphai+higherlogs} to compute $j_\alpha$, we collect all terms of order $\tau^{-3}(\ln\tau)^n$ to be
\begin{equation}
    \overset{3,n}{j_\alpha}
    =
        -e^2 \overset0A_\alpha \sum_{k=0}^n b_{0,k}^* b_{0,n-k}
        - ie\sum_{k=0}^n b_{0,k}^* \calD_\alpha b_{0,n-k}
        + \text{c.c.}
    ~.
\end{equation}
Using the expression \eqref{b0n} for $b_{0,n}$, we find that the two sums organize into simple binomial expansions that reduce to either $\delta_{n,0}$ and $\delta_{n,1}$:
\begin{equation}
    \overset{3,n}{j_\alpha}
    =
        \delta_{n,0}\left(
            - 2e^2 \overset0A_\alpha |b_0|^2
            - ie(b_0^*\calD_\alpha b_0 - b_0 \calD_\alpha b_0^*)
        \right)
        + 2\delta_{n,1} e^2|b_0|^2 \p_\alpha \overset1A_\tau
        + (b\to d^*)
    .
\end{equation}
Therefore as a corollary, it follows that
\begin{equation}\label{javanish}
    \overset{3,n}{j_\alpha}=0
    \qquad\text{ for all $n\geq 2$}
    .
\end{equation}

\bibliographystyle{jhep}
\bibliography{references}

\end{document}